\documentclass[trackchanges,twocolumn]{aastex701}
\usepackage{natbib}
\usepackage{float}
\usepackage{tikz}
\usepackage{bm}
\usepackage[percent]{overpic}
\usepackage{amsmath}
\usepackage{booktabs}

\usetikzlibrary{positioning, shapes.geometric, arrows.meta}

\hypersetup{linkcolor=red,citecolor=blue,filecolor=cyan,urlcolor=purple}

\begin{document}

\title{A Variability Metric Pipeline for AGN studies}

\correspondingauthor{Swayamtrupta Panda}
\email{swayamtrupta.panda@noirlab.edu}

\author[0009-0004-4993-5656]{Diego Martínez Collipal}
\affiliation{Astronomy Department, Universidad de Concepción, Casilla 160-C, Concepción 4030000, Chile}
\affiliation{Millennium Nucleus on Transversal Research and Technology to Explore Supermassive Black Holes (TITANs)}
\email{diego.collipal10@gmail.com}

\author[0000-0002-5854-7426]{Swayamtrupta Panda}\thanks{Gemini Science Fellow}
\affiliation{International Gemini Observatory/NSF NOIRLab, Casilla 603, La Serena, Chile}
\email{swayamtrupta.panda@noirlab.edu}

\begin{abstract}
Variability in active galactic nuclei (AGNs) is remarkably diverse, spanning multiple timescales, wavelengths, and physical processes. Variability probes accretion physics and the interplay between supermassive black holes and their surroundings. Building on early case studies, the field has shifted toward population-level analyses of increasingly large datasets, making scalable techniques crucial for upcoming time-domain surveys. We developed a Python-based pipeline to analyze AGN light curves using variability metrics and applied it to 14,061 nearby AGNs ($z\lesssim 0.09$) observed with ZTF. We examined metric distributions and correlations. Amplitude-based metrics (e.g., $F_{\rm var}$) help identify major flares but are sensitive to single deviant epochs. Temporal metrics such as $\eta$ provide a complementary check. Comparison with WISE infrared photometry suggests that $\eta$ traces the strength of AGN activity. Most Type~I and Type~II AGNs in our sample exhibit similar variability patterns based on these metrics. This overlap may partly reflect misclassifications that targeted spectroscopic follow-up could help resolve. To identify subsets of AGNs with distinctive variability, we used UMAP to group sources with similar metric values. The resulting embedding shows a region enriched in high-amplitude variables, including a subset with long-term flux trends. We find evidence that extremely variable quasars and changing-look AGNs may populate this region, which can help prioritize sources for follow-up. Our pipeline offers a scalable, survey-agnostic framework suitable for application to Vera C. Rubin Observatory's LSST and other ongoing and upcoming photometric surveys.
\end{abstract}

\section{Introduction}
\label{sec:intro}
Active galactic nuclei (AGN) display flux variations across the entire electromagnetic spectrum, from $\gamma$-rays to Radio, with typical timescales in the optical range spanning from several months to years \citep{Macleod2012}. Variability provides a unique way to study the sub-parsec scale of AGNs, which currently remains inaccessible through direct imaging. The typical timescales of the variability episodes impose direct constraints on the physical size of the emitting region, which helped to reveal that AGN luminosities originate in compact regions, establishing the connection with supermassive black holes \citep[SMBHs,][]{Ulrich1997}.\\

The characterization of the distinct variability types has improved our understanding of the physical processes in the vicinity of SMBHs. The discovery of extreme variability events such as Tidal disruption events \citep[TDEs,][]{rees1988, vanvelzen2020, Gezari2021} or Changing-look AGNs \citep[CL-AGNs,][]{lamassa2015discovery,ricci2023changing} has demonstrated that substantial changes in black hole activity may occur much faster and on relatively short timescales than previously thought \citep{Komossa2026}. Moreover, variability provides insights into the physics of the accretion process, with implications for SMBH growth; see \citet{paolillo2025} for a detailed review.\\

The study of AGN variability began as a collection of case studies. Early observations of sources such as NGC 4151 revealed the first evidence of brightness variations in AGNs \citep{Fitch1967}, and studies such as \citet{blandford1982} began to use variability as a tool to resolve AGN structure. These pioneering studies helped to establish variability as a defining property of AGNs, and it is now used as an efficient tool for identifying new AGN populations, improving our understanding of black hole demographics \citep{Villforth2010, lamassa2015discovery,Graham2015b,Baldassare2018ApJ,Hernandez-Garcia2025}. The development of dedicated long-term photometric campaigns gradually shifted the field from single-object studies to statistically meaningful populations. {A key survey in this transition was the Catalina Real-time Transient Survey \citep[CRTS,][]{drake2009}, designed to observe a total sky area of 26,000 deg$^{2}$ in the optical range. CRTS operated as a public optical transient survey from 2007 to 2019, providing approximately 12 years of data \citep{Drake2025}. This survey helped to characterize extreme events, such as major AGN flares \citep{Graham2017} or changing state episodes \citep{graham2020}. CRTS observations from the 0.7-m Catalina Schmidt telescope were obtained without a filter using a single 4k$\times$4k CCD, with a pixel scale of 2.5 arcsec\,pixel$^{-1}$ and a field of view of approximately 8 deg$^{2}$. Although this design was crucial to maximize sky coverage, it also prevented CRTS from exploring the wavelength dependence of variability. This limitation was assessed by subsequent campaigns, such as the Panoramic Survey Telescope and Rapid Response System \citep[Pan-STARRS,][]{Chambers2016}. The Pan-STARRS1 (PS1) $3\pi$ survey repeatedly imaged approximately 30,000 deg$^{2}$ of the sky north of $\delta=-30^\circ$ during 2010-2014, using five broadband filters ($g_{\mathrm{P1}}$, $r_{\mathrm{P1}}$, $i_{\mathrm{P1}}$, $z_{\mathrm{P1}}$, and $y_{\mathrm{P1}}$) spanning approximately $4000$--$11{,}000\,\AA$. Pan-STARRS provided multi-epoch, multi-band photometry for millions of objects, leading to further improvements in AGN selection methods \citep{Morganson2014} and the study of the spectral nature of variability. A relevant result delivered by this survey was the confirmation that AGN variability amplitudes are generally larger at shorter wavelengths \citep{Morganson2014,hung2016}. These and other photometric campaigns, e.g., the Sloan Digital Sky Survey \citep[SDSS,][]{Sesar2007}, the All-Sky Automated Survey for Supernovae \citep[ASAS-SN,][]{Shapee2014}, the Kepler mission \citep[Kepler,][]{Borucki2010}, the Transiting Exoplanet Survey Satellite \citep[TESS,][]{Ricker2014}, the European Space Agency’s (ESA) Gaia mission \citep[Gaia,][]{Gaia2016}}, inspired the design of modern time-domain surveys. Most notably, the Zwicky Transient Facility \citep[ZTF,][]{Bellm2019,graham2019} represents the current state-of-the-art time-domain survey \citep{graham2020}. Since March 2018, ZTF has been monitoring the northern hemisphere with a 576-megapixel camera \citep{dekany2016}, which covers a $47 deg^{2}$ field of view. The survey observes in three optical bands (ZTF-r, ZTF-g, and ZTF-i), with a high cadence of measurements every 2-3 nights. ZTF design combines significant sky coverage with multi-band photometry and high observational cadence, providing densely sampled light curves suitable for AGN studies on multiple timescales. In addition, ZTF includes a real-time alert system \citep{patterson2019}, allowing the rapid identification and follow-up of transient events.\\

The development of time-domain surveys has enabled the characterization of variability for millions of AGNs, providing a much deeper understanding of the underlying physical phenomena, and revealing a striking diversity of variability behaviors. AGNs commonly show non-periodic, stochastic brightness fluctuations. This type of variability is likely driven by thermal instabilities within the accretion flow, which in some cases can be mathematically described as a damped random walk  \citep[DRW,][]{Kelly2009}. However, multiple objects deviate from this behavior \citep{kozlowski2016,paolillo2025}. Other AGNs also show significantly larger changes in their brightness, defining the class of extreme-variability quasars \citep[EVQs,][]{rumbaugh2018}. A subset of these extreme objects also change their spectral type between Type I and Type II. These objects are known as changing-look AGNs and are thought to be produced by obscuration events or changes in the accretion activity \citep{lamassa2015discovery,graham2020,Macleod2012}. Another distinct type of extreme variability is produced by the tidal disruption of a star passing close to the SMBH, i.e., a tidal disruption event (TDE), which produces large transient flux increments \citep{rees1988,vanvelzen2020,Gezari2021}. More recently, the development of real-time alert systems has helped to detect the turn-on of AGNs in non-active galaxies, a new class of objects called newborn AGNs \citep{Arevalo2024,Sanchez-Saez2024,Hernandez-Garcia2025}.\\

In this context, AGN studies must develop bulk-applicable techniques to characterize variability. Variability metrics are a natural approach to this necessity. Metrics are simple statistical descriptors that can be extracted from a light curve. {An example is the $F_{\rm var}$ metric, which has been widely used in AGN studies as a robust way to quantify the amplitude/scatter of the variability, accounting for observational noise \citep{Edelson2002,Peterson2004,benati2025}. Another case is the Stetson index \textit{J} \citep{Welch1993}, which assesses the temporal coherence of the fluctuations, along with their significance against the observational noise. This metric is commonly used to detect significantly variable sources \citep{graham2014,kinemuchi2020,Ma2024}. Metrics have been successfully applied in multiple stellar and AGN studies \citep{sokolovsky2017, kim2011, arevalo2026}. They are directly obtained from the flux distribution of a light curve and are model-independent, since they do not assume any physical mechanism for the variability. Therefore, they are ideal for large-scale studies of heterogeneous samples. Moreover, variability metrics provide a quantitative description of a light curve, which can serve as a benchmark to characterize large populations and potentially identify promising targets for spectro-photometric follow-up observations, as successfully done by previous works \citep{Richards2011,macleod2019changing,graham2020,sanchez-saez2021,Hon2022,Lopez-Navas2022}}.\\

In this work, we present an open-source Python-based pipeline designed to compute multiple variability metrics for a given light curve, allowing us to record distinct variations based on their temporal flux change. We catalog these variability metrics for a large sample of AGN light curves from ZTF. We explore correlations among the set of metrics and provide a catalog of metric values for the objects in our sample, establishing a benchmark for future variability studies with the pipeline. We assess the use of metrics to select promising science cases for spectroscopic follow-up. An efficient selection of targets from large photometric samples represents a powerful approach to reduce the required observational resources. Finally, we explore the use of metrics for clustering analysis with UMAP, and we discuss the impact of baseline, host galaxy contamination, and sample properties in our results.\\ 

This work is organized as follows. In Section \ref{sec:data}, we present our studied sample and the processing of the photometric data. In Section \ref{subsec:metric_dist}, we analyzed the resulting metric distributions and inspected outliers. Section \ref{subsec:correlations} presents the correlations between different metrics. Section explores the differences between Type I and Type II AGNs in terms of metrics \ref{subsec:type1_type2}. Section \ref{subsec:umap} presents the results of the UMAP analysis. Finally, we discuss the caveats of our results in Section \ref{sec:discussion}, and summarize the main findings in Section \ref{sec:summary}.

\section{Variability metrics}
In the present work, we collected multiple variability metrics from previous studies. These particular metrics were selected based on their definition, as they may potentially be useful to characterize AGN variability. We focused on metrics that evaluate the amplitude and temporal coherence of the variability, as these are key characteristics of stochastic variability. We also included metrics designed to detect outliers in the flux distribution, as well as metrics to detect long-term variability trends. These last metrics were selected for the efficient detection of sources showing extreme variations or unusual behaviors, such as EVQs, CLAGNs, and TDEs. Our complete set of metrics is described in Appendix \ref{app:metrics}.

\section{Data} \label{sec:data}

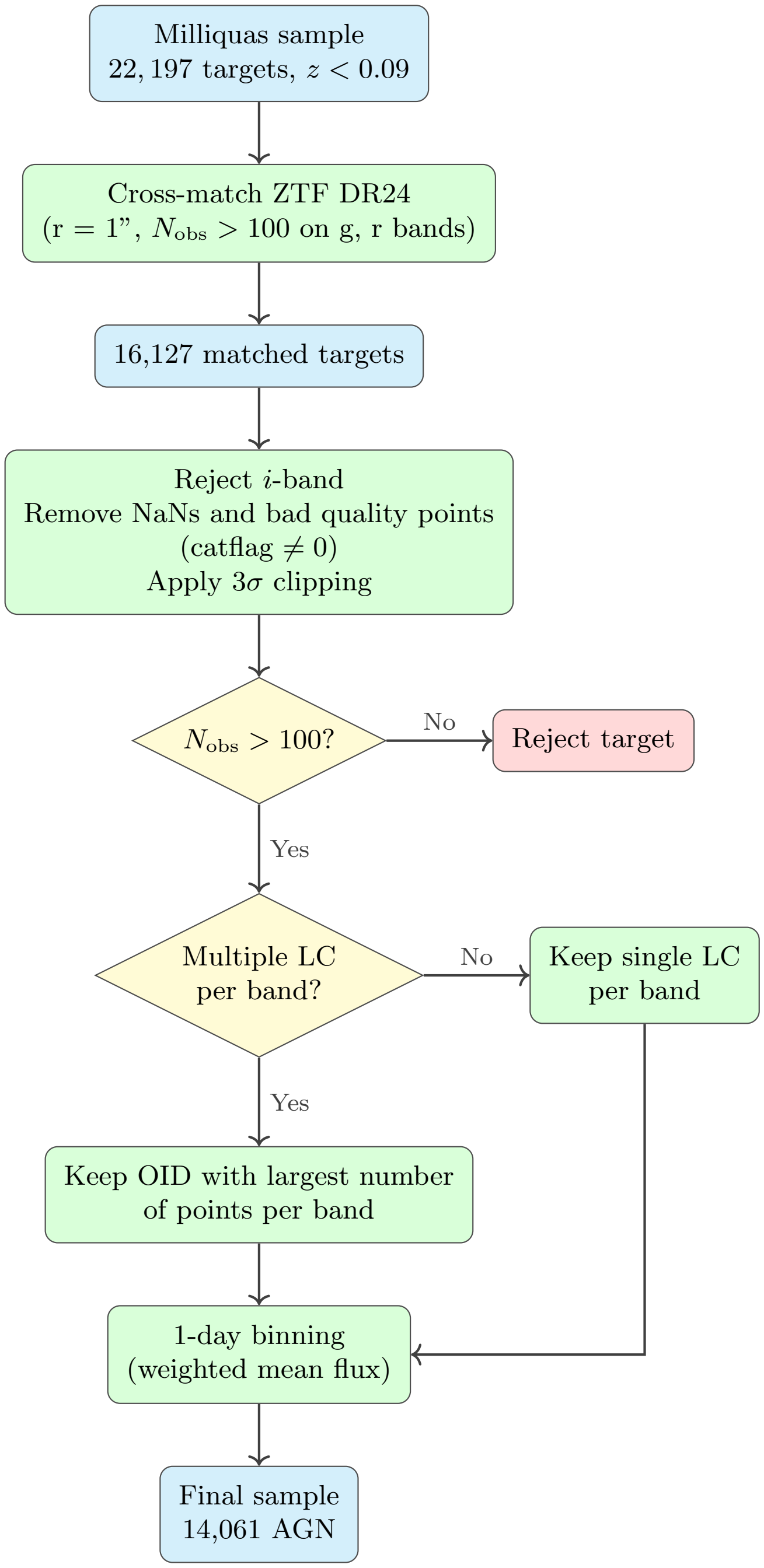
\begin{figure}
\centering
\begin{tikzpicture}[
    node distance=0.7cm,
    every node/.style={font=\small},
    process/.style={
        rectangle,
        rounded corners,
        draw=black!70,
        fill=cyan!15,        
        align=center,
        inner sep=6pt     
    },
    filter/.style={
        rectangle,
        rounded corners,
        draw=black!70,
        fill=green!15,      
        align=center,
        inner sep=6pt
    },
    reject/.style={
        rectangle,
        rounded corners,
        draw=black!70,
        fill=red!15,         
        align=center,
        inner sep=6pt
    },
    decision/.style={
        diamond,
        draw=black!70,
        fill=yellow!20,    
        aspect=2,
        align=center,
        inner sep=3pt
    },
    arrow/.style={
        ->,
        thick,
        darkgray
    }
]

\node[process] (milliquas)
{Milliquas sample \\ $22,197$ targets, $z<0.09$};

\node[filter, below=of milliquas] (ztf)
{Cross-match ZTF DR24\\ (r = 1'', $N_{\rm obs} > 100$ on g, r bands)};

\node[process, below=of ztf] (matched)
{16,127 matched targets};

\node[filter, below=of matched] (clean)
{Reject $i$-band \\ Remove NaNs and bad quality points\\ (catflag $\neq 0$)\\ Apply $3\sigma$ clipping};

\node[decision, below=of clean] (nobs)
{$N_{\rm obs} > 100$?};

\node[decision, below=1.0cm of nobs] (multi)
{Multiple LC\\ per band?};

\node[filter, below=1.0cm of multi] (select)
{Keep OID with largest number\\ of points per band};

\node[filter, below=of select] (rebin)
{1-day binning\\ (weighted mean flux)};

\node[process, below=of rebin] (final)
{Final sample\\ 14,061 AGN};

\node[reject, right=1.2cm of nobs] (reject1)
{Reject target};

\node[filter, right=1.2cm of multi] (keepall)
{Keep single LC\\ per band};

\draw[arrow] (milliquas) -- (ztf);
\draw[arrow] (ztf) -- (matched);
\draw[arrow] (matched) -- (clean);
\draw[arrow] (clean) -- (nobs);

\draw[arrow] (nobs) -- node[right, font=\footnotesize] {Yes} (multi);
\draw[arrow] (nobs.east) -- node[above, font=\footnotesize] {No} (reject1.west);

\draw[arrow] (multi) -- node[right, font=\footnotesize] {Yes} (select);
\draw[arrow] (multi.east) -- node[above, font=\footnotesize] {No} (keepall.west);

\draw[arrow] (keepall) |- (rebin);
\draw[arrow] (select) -- (rebin);
\draw[arrow] (rebin) -- (final);

\end{tikzpicture}
\caption{Workflow used to process the ZTF light curves.}
\label{fig:flowchart}
\end{figure}
In the present work, we restricted the analysis to nearby AGNs $( z < 0.09 )$. With this cut, we expect to avoid mixing objects at different redshift ranges, as their evolutionary states may differ. At the same time, the low redshift of our sample minimizes the effect of cosmic dilatation, making the timescales measured in the observed-frame directly comparable to the AGN rest-frame, and ensures that the ZTF photometric bands are centered on the strong emission lines of H$\beta$ and H$\alpha$ for all objects, minimizing wavelength-dependent variations in the light curves. We discuss more about the potential redshift effects in our results in Section \ref{subsec:redshift_range} \\

To collect our initial sample, we applied this redshift cut to The Million Quasars Catalog \citep[Milliquas,][]{Flesch2023}, finding a total of 22,197 potential targets. We searched photometric data for these targets in the Zwicky Transient Facility Data Release 24 (ZTF DR24). For this, we did URL-based queries to the IRSA archive\footnote{\url{https://irsa.ipac.caltech.edu/Missions/ztf.html}} centered on the target's coordinates with a radius of 1 arcsec. This radius limits the positional offset of the matched ZTF sources. We emphasize that the retrieved photometry can include host-galaxy light blended with the nucleus at the ZTF spatial resolution. We discuss more about the host galaxy contamination and its impact on the variability trends and metrics performance in Section \ref{subsec:redshift_range}.

Additionally, we imposed a minimum of 100 observed data points for each target on each ZTF band to ensure enough sampling for our metrics analysis. Using these constraints, we found a total of 16,127 matched targets. \\

\begin{figure}
\centering
\includegraphics[width=\columnwidth]{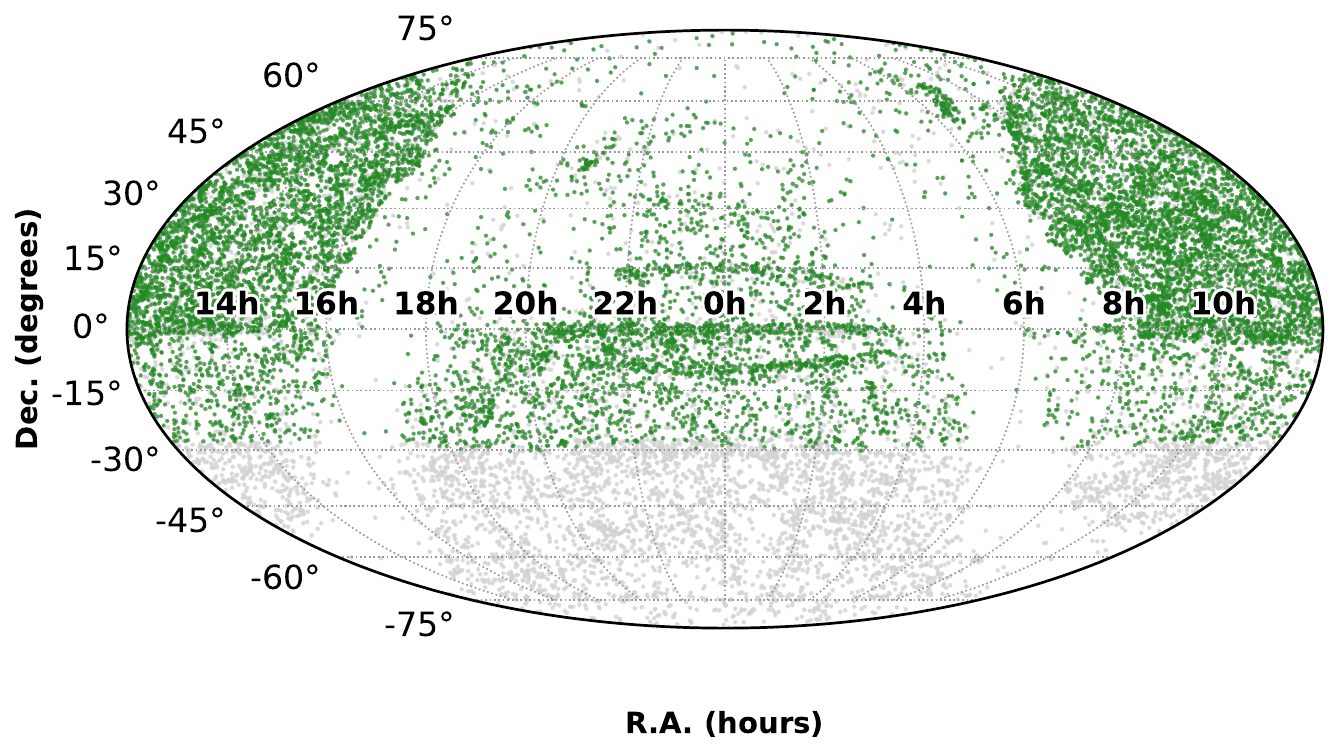}
\caption{Sky distribution of our sample. The initial Milliquas sample is presented in grey, while our final parent sample after constraints is colored green.}
\label{fig:skymap}
\end{figure}

We processed the resulting data with the following steps:
\begin{itemize}
    \item For each target, we organized the multiple ZTF light curves according to their object ID (oid), and we only kept observations in the ZTF g- and r-bands, as the i-band observations were found to be scarce and sparse.
    \item For the resulting light curves, we cleaned NaN values and bad-quality observations using the ZTF quality flag ``catflag''. We converted the ZTF AB magnitudes to physical fluxes in cgs units (erg cm$^{-2}$ s$^{-1}$ Hz$^{-1}$) using:
    \begin{equation}
        F_{i}=10^{-0.4(m_{i}+48.6)}
    \end{equation} 
    where the factor 48.6 corresponds to the zero point. We used the derived fluxes to apply a 3$\sigma$ clipping to remove outliers. After this cleaning process, we tested whether the resulting light curves contained more than 100 observations; those with fewer were rejected. Finally, we rebinned all light curves to 1-day bin fluxes, where the flux for each bin corresponds to the weighted average of the observations within that particular bin.
    \item We required each target to have at least one light curve in ZTF g and r bands. In some cases, we found multiple light curves for a given ZTF field and filter. In those cases, we kept the observations with the highest number of data points.    
\end{itemize}
The procedure is summarized in Figure \ref{fig:flowchart}. Our final sample consists of 14,061 objects, and we present its sky distribution in Figure \ref{fig:skymap}. We computed all the metrics for these objects, both in the g- and r-bands. For this, we first normalized each light curve by its mean flux to make the results comparable between objects with different luminosities. This standardization also helps with the dimensionality reduction analyses presented in Section \ref{subsec:umap}.

\subsection{Additional WISE data}
We additionally included photometric data from the Wide-field Infrared Survey Explorer (WISE) \citep{Wright2010} for our analysis. For this, we cross-matched the objects in our final sample with the WISE-ALL-Sky Source Catalog \citep{Cutri2012}, using the same search radius of 1 arcsec. We got a total of 13,607 unique matches, which correspond to $\sim 98\%$ of the initial sample. We retrieved the corresponding WISE magnitudes W1, W2, W3, and W4 for these objects, which are measured with infrared filters centered at 3.4, 4.6, 12, and 22 $\mu m$ \citep{Wright2010}. 


\section{Results} \label{sec:results}
\subsection{Metrics distributions} \label{subsec:metric_dist}
We analyze the distribution of values for each metric. All distributions and corresponding statistics are provided in the appendix \ref{app:metric_distributions}.

\subsubsection{Flux distribution metrics}
We firstly identified that the distribution of the $g_{1}$ metric \footnote{$g_{1}$ is the third standardized moment of the flux distribution, i.e., the skewness} is clearly shifted towards positive values ($\mu \sim 1.61\,, \sigma\sim 0.67$), indicating that a positive asymmetry is a common property of the flux distribution in our light curves. Moreover, the kurtosis of this distribution is among the lowest in the dataset, which suggests that extreme events are rarely reflected in this metric.\\

The distributions for the percentile ratios Mid20, Mid35, Mid50, Mid65, and Mid80 are concentrated around average values of 0.113, 0.206, 0.316, 0.463, and 0.698, respectively. With very low standard deviations in the order of $\sim 10^{-3}$. The average values are systematically below the expected values for a Gaussian distribution, which are Mid20 $\approx  0.154$, Mid35 $\approx  0.276$, Mid50 $\approx 0.410$, Mid65 $\approx 0.568$, and Mid80 $\approx  0.779$. The lower values in the percentage ratios mean that the flux distribution for AGN is less concentrated around the center than a Gaussian. In addition, the observed low dispersion suggests that this is also a common property across AGNs in our sample. The analysis of the kurtosis and skewness of the distributions shows that Mid20 has a positive asymmetry ($g_{1} = 1.79$) and high kurtosis ($g_{2} = 8.06$), indicating that a reduced number of sources possess a substantially broader central region than the rest of the sample. These skewness and kurtosis values gradually decreased from Mid20 to Mid80. Therefore, our AGNs show a higher degree of diversity in the innermost parts of the flux distribution than in their overall shape.

\subsubsection{Amplitude/scatter metrics}
\begin{figure}
    \centering
    \includegraphics[width=\columnwidth]{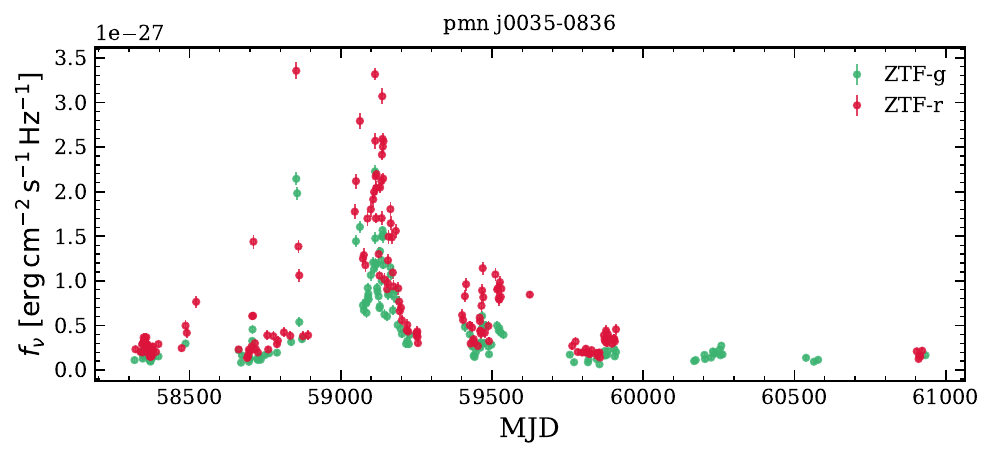}
    \caption{Outlier source identified from the $F_{var}$ distribution.}
    \label{fig:outlier_fvar}
\end{figure}

\begin{figure}
    \centering
    \includegraphics[width=\columnwidth]{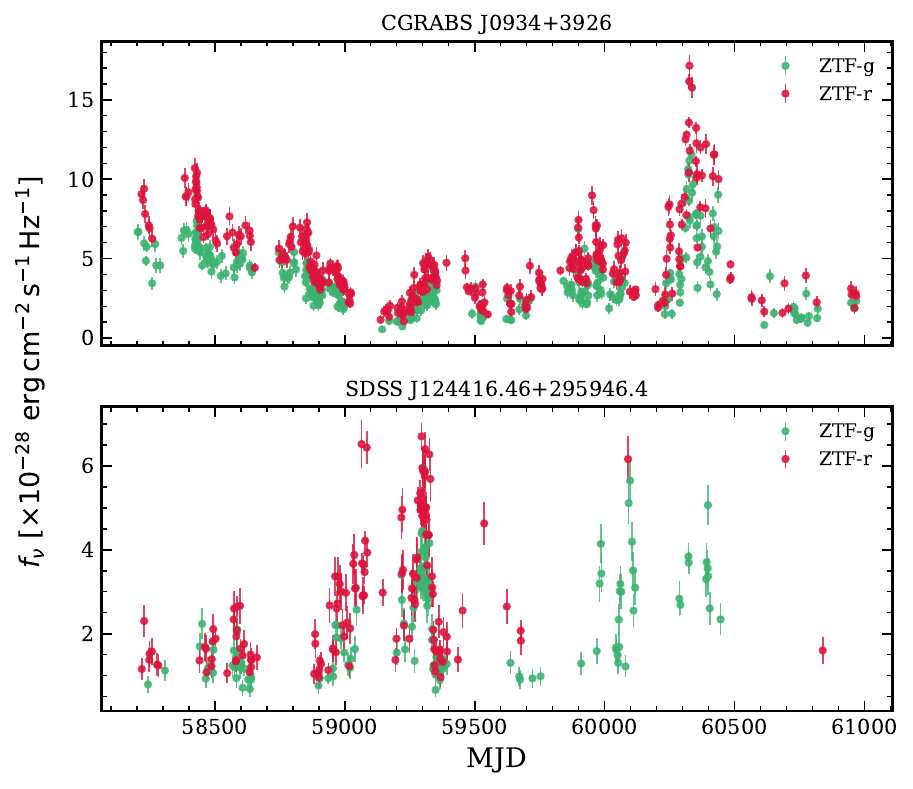}
    \caption{Outlier sources identified from the MAD distribution.}
    \label{fig:outlier_MAD}
\end{figure}

The distributions for all scatter metrics (A, $\sigma_{w}$, MAD, RoMS, $F_{\rm var}$, IQR, BBA) \footnote{\textbf{A}: light curve amplitude, $\mathbf{\sigma_{w}}$: weighted standard deviation, \textbf{MAD}: median absolute deviation from the median flux, \textbf{RoMs}: robust median statistic (Eq. \ref{equation:ROMS}), $\mathbf{F_{var}}$: fractional variability amplitude (Eq. \ref{equation:FVAR}), \textbf{IQR}: interquartile range of the flux distribution, \textbf{BBA}: flux amplitude measured in the Bayesian block representation of the light curve.} tend to show extreme outliers and positive asymmetries. An extreme degree of scatter could be produced either by extreme single flare events or by stochastic fluctuations of high amplitude. We explored the nature of our outliers by visual inspection of extreme values in the scatter metrics. We present one such case in Figure \ref{fig:outlier_fvar}. The object identified as PMN J0035-0836 has the highest $F_{\rm var}$ value in our sample ($F_{\rm var} = 1.93$ in the g-band), which is around 20 times the median value of the distribution. This high value of $F_{\rm var}$ was likely produced by the single flare event observed around MJD = 59000. The event shows a flux increment of around 3 times in the r-band, which was followed by a fast decline back to a lower state that is present during the rest of the monitoring. We have found that this event is also an outlier in other scatter metrics. In this case, the event also has the highest values of A, $\sigma_{w}$, IQR, and BBA. PMN J0035-0836 is the radio counterpart of the Fermi gamma-ray source 4FGL J0035.8-0837, cataloged as a blazar candidate of uncertain type (BCU) \citep{Foschini2022}. We interestingly noticed that \citet{Foschini2022} reported a redshift of z=0.965 for this object, while Milliquas reports a z=0.032. Given the blazar classification of PMN J0035-0836, the detected flare may have its origin in the jet emission. Interestingly, gamma-ray emission in the 3-10 GeV band was also observed around July-August 2020 (MJD $\sim$ 59000) in the object coordinates \citep{Chen2024}, suggesting a potential multi-wavelength event.\\
\begin{figure}[hb]
    \centering
    \includegraphics[width=\columnwidth]{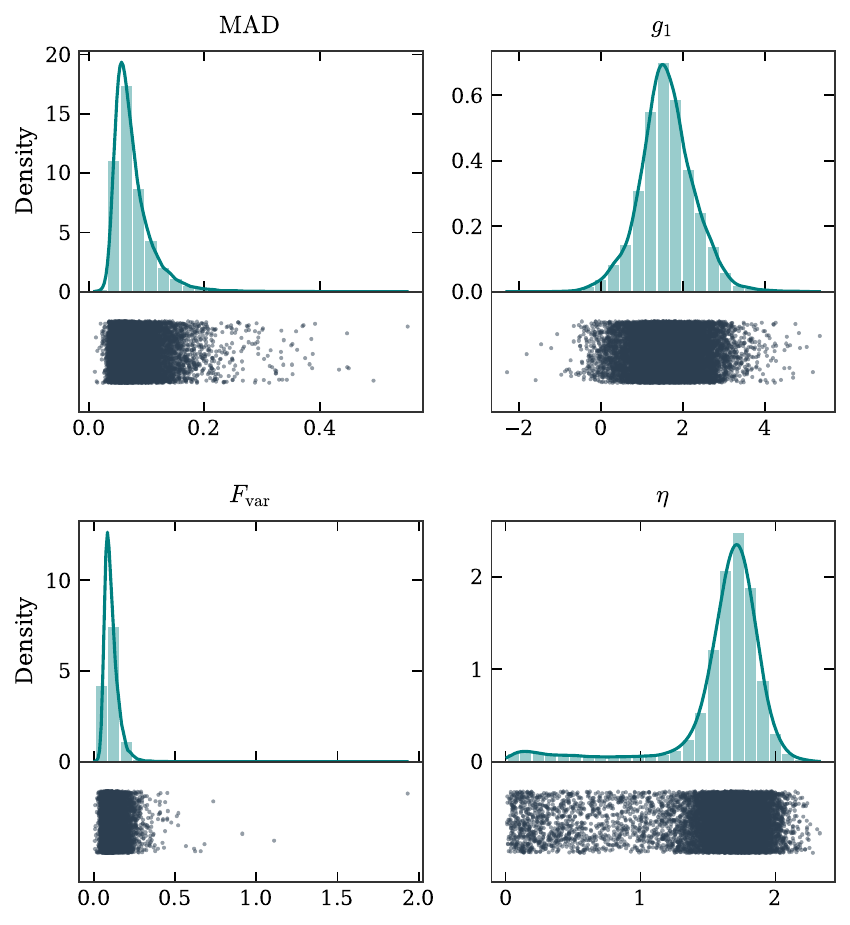}
    \caption{Metric distributions discussed in the text: The top of each panel shows the probability density histograms (bars) and the kernel density estimate (KDE, solid curve). The bottom shows the actual metric values for each source dispersed along the x-axis with a jitter for visualization. These distributions correspond to the $g$-band light curves. The complete set of distributions can be found in Fig. \ref{fig:metric_dist_all}.}
    \label{fig:metric_dist}
\end{figure}
It is interesting to note that MAD is the only scatter metric that does not identify this event as the most extreme one. As shown in Equation \ref{equation:MAD}, the MAD metric measures the flux difference with respect to the median, therefore being intrinsically robust to outliers. Since PMN J0035-0836 is in a steady state during most of the observation, its brief high-flux state has a reduced impact on the final value of MAD, explaining the difference with respect to other metrics. The distinct nature of MAD is likely reflected in the distribution of points in the MAD panel of Figure \ref{fig:metric_dist}, which shows that multiple objects share a similar degree of deviation in this metric, instead of a single outstanding outlier as observed in the distribution of $F_{var}$ or A. Figure \ref{fig:outlier_MAD} shows the two sources with the highest MAD values in our sample. 
Both sources show flare-like behavior. CGRaBS J0934+3926 presents a significant flux increment around MJD=60500, followed by a quiescent state. In contrast, the characterization of SDSS J124416.46+295946.4 is less clear due to its sparse light curve. CGRaBS J0934+3926 is a blazar associated with the Fermi source 4FGL J0934.3+3926 \citep{abdollahi2020}. The observed flare may have its origin in the jet emission in this case as well. Interestingly, previous studies have classified this object as a BL Lac \citep{Plotkin2008}, but \citet{Liang2025} recently proposed its reclassification to a flat-spectrum radio quasar (FSRQ) due to the detection of emission lines in its spectra. This study also obtained a new spectroscopic redshift of z=1.1691 for this object, while Milliquas reports a redshift of z=0.044. We note that this second discrepancy suggests that some higher redshift objects could be present in our sample due to redshift miscalculation. \\

The nature of the event in SDSS J124416.46+295946.4 is less clear. This object has a potential infrared counterpart identified as WISEA J124416.45+295946.2 at 0.24 arcsec from the object coordinates, and two potential radio counterparts, FIRST J124416.4+295946 and NVSS J124416+295947, identified in NED \footnote{NASA/IPAC Extragalactic Database \url{https://ned.ipac.caltech.edu/}}. This object is identified as a Radio source in Milliquas, but its archival SDSS BOSS spectra \citep{dawson2013} show no detectable AGN activity \footnote{SDSS spectra: \url{https://skyserver.sdss.org/dr20/VisualTools/explore/summary?ra=191.068583&dec=29.996233}}. Therefore, the classification of this object needs to be studied further in order to interpret the observed flare.\\

We have seen through these examples that, despite MAD being a metric intrinsically robust to flux outliers, it remains sensitive and useful for detecting extreme variability events.

\subsubsection{The $\eta$ metric}
\begin{figure}
    \centering
\includegraphics[width=\columnwidth]{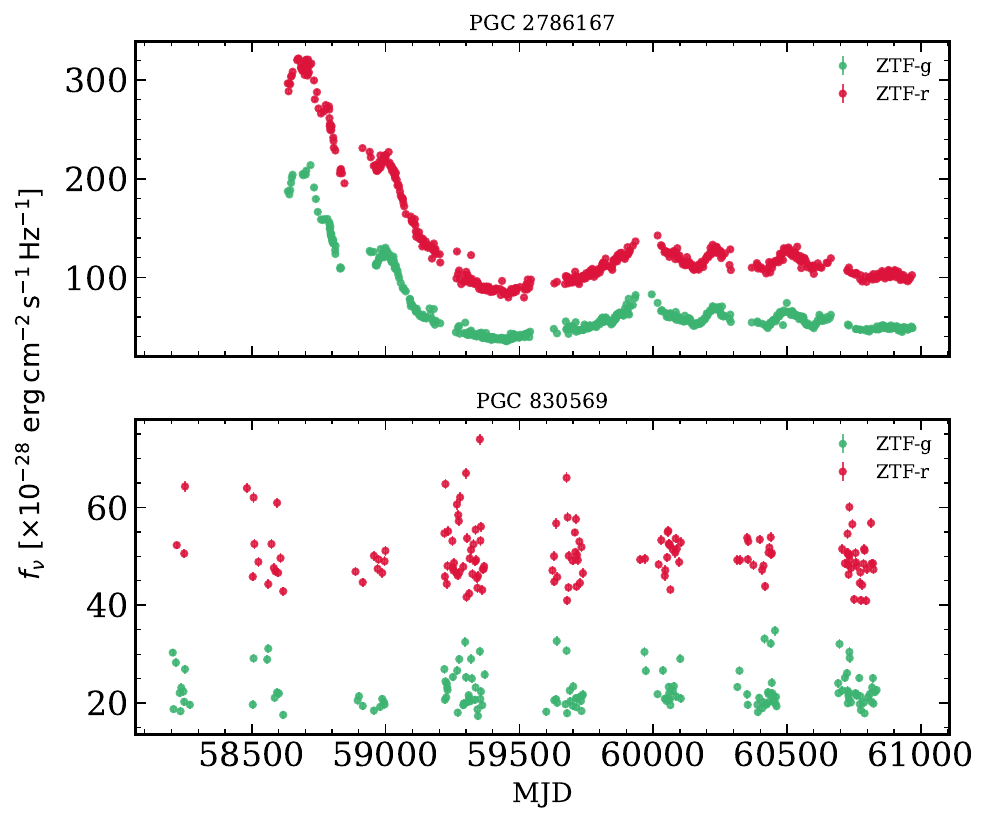}
    \caption{Representative cases in the $\eta$ distribution. The top panel shows the source with the lowest $\eta$ value ($\eta \approx 0.008$). Whereas the lower panel shows the source with the highest $\eta$ value ($\eta \approx 2.37$). Values correspond to the g-band.}
    \label{fig:outlier_eta}
\end{figure}

The distribution of the $\eta$ metric is particularly different from the rest. The $\eta$ metric is constructed from the differences between subsequent flux measurements and normalized by the total flux variance (Eq. \ref{eq:eta}). This metric is one of the few in our set that evaluates the temporal coherence of the variability. Its distribution shows a strong negative asymmetry that seems driven by a distinct population of objects rather than a few outliers (see the $\eta$ panel in Figure \ref{fig:metric_dist}). Small values of $\eta$ indicate a high degree of temporal coherence between subsequent flux measurements, whereas high values indicate more disconnected episodes. Both behaviors seem to be present in our sample. We illustrate the potential of $\eta$ to differentiate these behaviors in Figure \ref{fig:outlier_eta}, where we compared the light curves of objects at the extremes of the $\eta$ distribution.

In the upper panel, PGC 2786167 displays a smooth flux decline to a steady state. In the lower panel, PGC 830569 shows unclear behavior and sparse flux measurements. The difference is clearly exaggerated by the much better sampling in the light curve of PGC 2786167. In that sense, the $\eta$ metric can also be used to reject poorly sampled light curves. The low-$\eta$ object PGC 2786167 is also cataloged as ICRF J211401.1+820448, S5 2116+81, 2116+818, and JVAS J2114+8204. Milliquas classifies this object as Type I, consistent with previous studies, which also detected radio-loud activity \citep{bassani2016,molina2008}. On the other hand, the high-$\eta$ case PGC 830569 is also known as 6dFGS gJ122105.3-211534 or LEDA 830569. Although Milliquas classifies this object as Type I, the available spectra from the 6dF Galaxy Survey \citep{jones2009} shows a host-dominated emission with no obvious broad balmer lines, consistent with a Type 2 \footnote{6dFGS spectra: \url{http://www-wfau.roe.ac.uk/6dFGS/cgi-bin/show.cgi?release=dr3&targetname=g1221053-211534&tid=-1&specid=59158&ra=185.27225&dec=-21.2595}}. It is possible that the AGN activity in this object is weak. Later in the text, we further explore a potential relation between $\eta$ and the nuclear activity (Sec. \ref{subsec:type1_type2}).   

\subsubsection{The Linear metric}
\begin{figure}
    \centering
\includegraphics[width=\columnwidth]{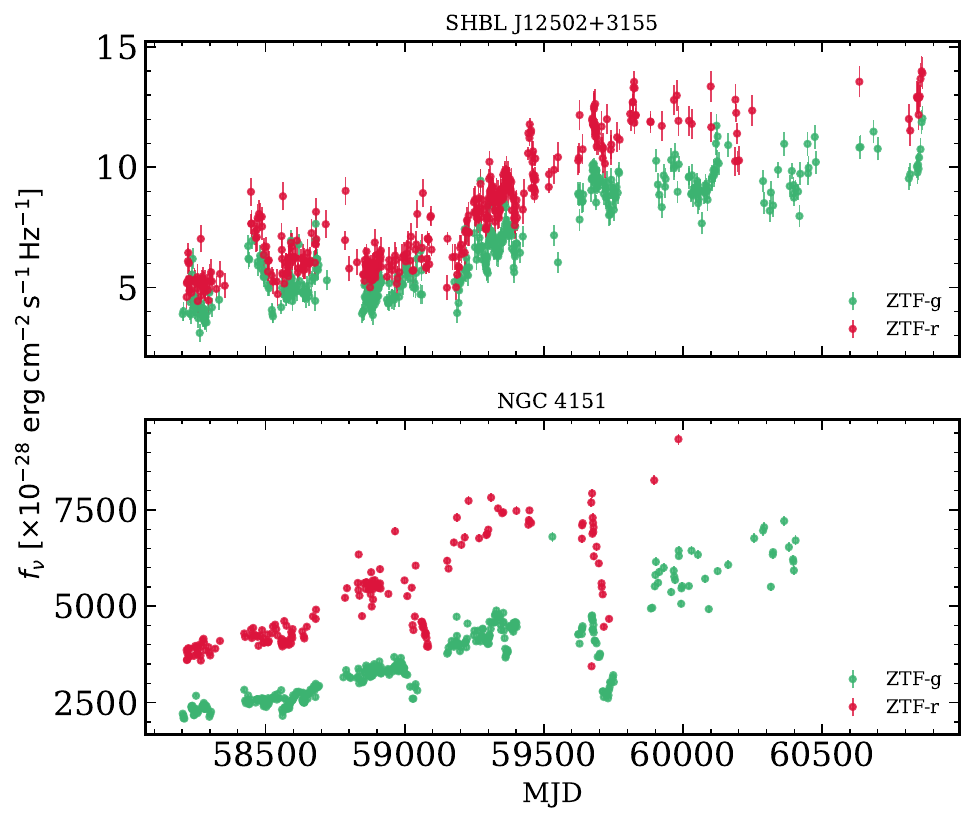}
    \caption{Extreme events in the Linear distribution.}
    \label{fig:outlier_linear}
\end{figure}

The Linear metric represents a potential tool to identify monotonically increasing or decreasing trends, which could be indicative of major changes in the accretion activity. We test this hypothesis by analyzing extreme outliers in the Linear distribution. We found two interesting objects, presented in Figure \ref{fig:outlier_linear}. In the lower panel, the Seyfert galaxy NGC 4151 is a well-known changing-look AGN that has undergone repeated optical spectral-state changes since the 1980s, including episodes in which broad $\mathrm{H}\beta$ disappeared and later re-emerged \citep{penston1984,shapovalova2008}. The monotonic flux increment observed in our ZTF epochs is consistent with NGC 4151 entering a new active phase, as described in \citet{feng2024}. The upper panel object, SHBL J12502+3155, is also known as RX J1250.2+3156 and is associated with the Fermi-LAT source 4FGL J1250.2+3157, which is classified as a BL Lac object in the Fermi-LAT catalog \citep{abdollahi2020}. The BL Lac classification suggests that the optical emission may be dominated by the relativistic jet, but we were unable to find more information about this target. The persistent brightening reported here represents an interesting case for further study.\\

The presented analyses highlight the potential of our metrics catalog to unveil interesting sources for follow-up observations.  
\subsection{Correlations}\label{subsec:correlations}
\begin{figure*}
\centering
\includegraphics[scale=0.6]{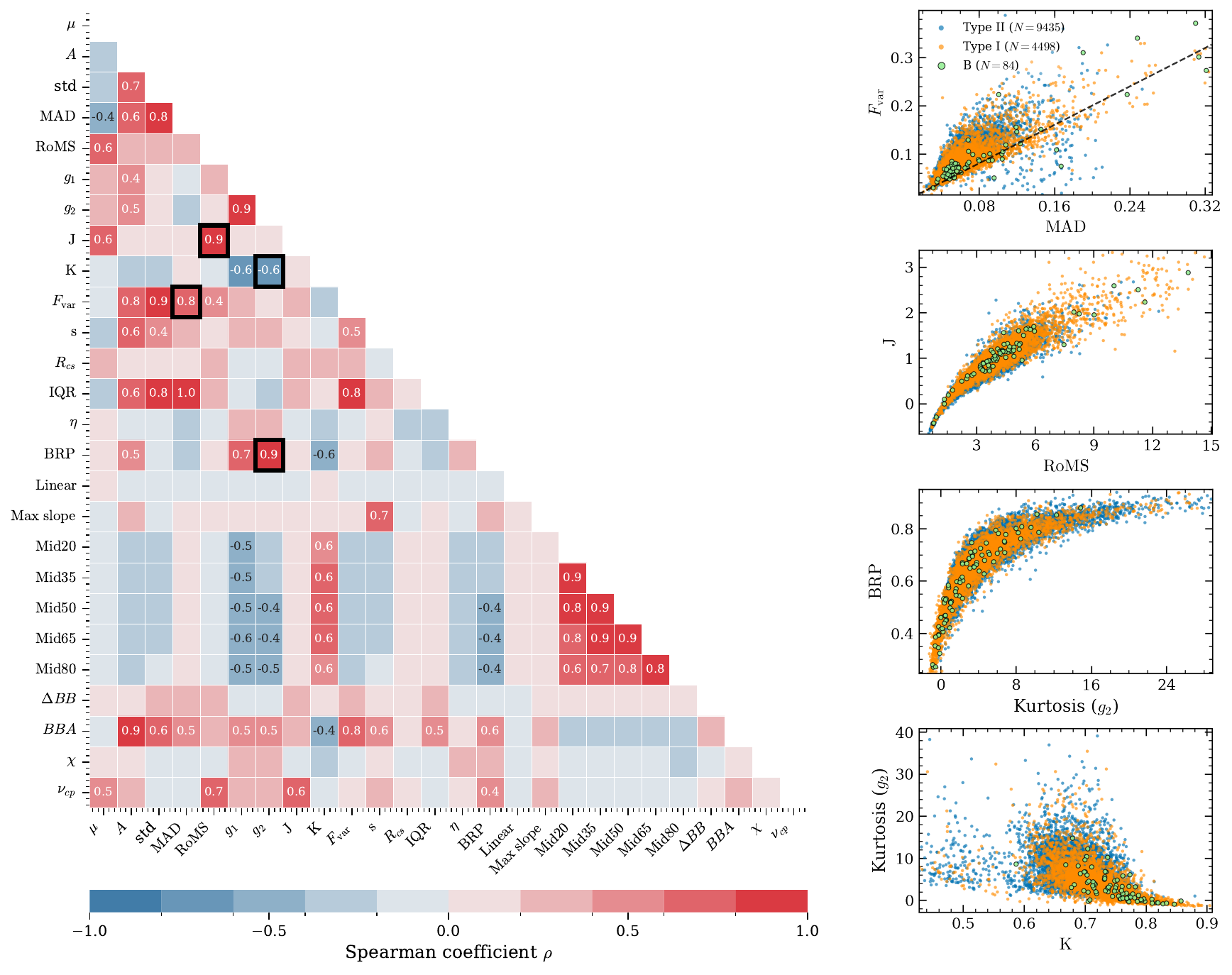}
    \caption{Left: Correlation matrix for the calculated metrics in the g-band. The colorbar shows the Spearman correlation coefficient between each pair. For significant correlations $(\rho_{s} > 0.4 )$, we show the exact value of $\rho_{s}$ inside the matrix element. Matrix elements corresponding to the scatter plots on the right are highlighted with black contours. Right: Scatter plots for interesting metric pairs. Objects are colored according to their classification in the Milliquas catalog; blue for Type II AGNs, orange for Type I, and green for Blazars. For the upper panel, we also included the expected 1:1 correlation between $\mathrm{MAD}$ and $F_{\rm var}$ with a black dashed line.}
\label{fig:corrmatrix}
\end{figure*}

Multiple metrics in our set are expected to be correlated, either because they are different ways to quantify the same thing, such as A and BBA, or because they could follow similar trends (both $g_{1}$ and $g_{2}$ would increase in objects with flare-like behavior). We studied these correlations by computing the Spearman correlation coefficient ($\rho_{s}$) between each pair of metrics. The results are presented in a correlation matrix in Figure \ref{fig:corrmatrix}.\\

We observed strong positive correlations between the scatter metrics (A, BBA, $\sigma$, MAD, $F_{\rm var}$, and IQR). Although the correlation coefficients are strong, some cases show a noticeable degree of scatter. An example of this is shown in the $F_{\rm var}$ vs MAD plane (top-right panel in Figure \ref{fig:corrmatrix}), where we observed a high dispersion and a clear offset with respect to the 1:1 relation. This offset suggests that large $F_{\rm var}$ values do not correspond to equally high values in MAD, as previously identified in the object of Figure \ref{fig:outlier_fvar}.  {The robust median statistic (RoMS) correlates weakly with other scatter metrics. This is probably related to the fact that RoMS quantifies the deviations relative to the photometric uncertainties (Eq. \ref{equation:ROMS}). Indeed, the metric shows a significant correlation with $F_{\rm var}$, which is the other scatter metric that includes the photometric uncertainties in its definition. RoMS also shows strong correlations with J and, to a smaller degree, with $\nu_{cp}$. These three metrics significantly correlate with the object flux scale measured by $\mu$. A natural reason for this is that both RoMS and J measure the deviations with respect to the flux uncertainties. Uncertainties are expected to be lower for brighter objects, affecting the value of these metrics. The case for $\nu_{cp}$ is slightly different. This metric is the ratio between the number of blocks in the Bayesian block representation of a light curve and the total baseline. The Bayesian block algorithm determines the optimal segmentation of a light curve using the fitness function, which employs the flux uncertainties to weight the measurements. Flux Measurements with lower errors represent stronger evidence of a flux change, which can result in an additional block. In this way, brighter objects will show smaller uncertainties, which influences the number of blocks determined by the algorithm. For further details, see \citet{scargle2013}, and the Astropy Bayesian block documentation\footnote{\url{https://docs.astropy.org/en/stable/api/astropy.stats.bayesian_blocks.html}}.} \\

\begin{figure}
\centering
\includegraphics[width=\columnwidth]{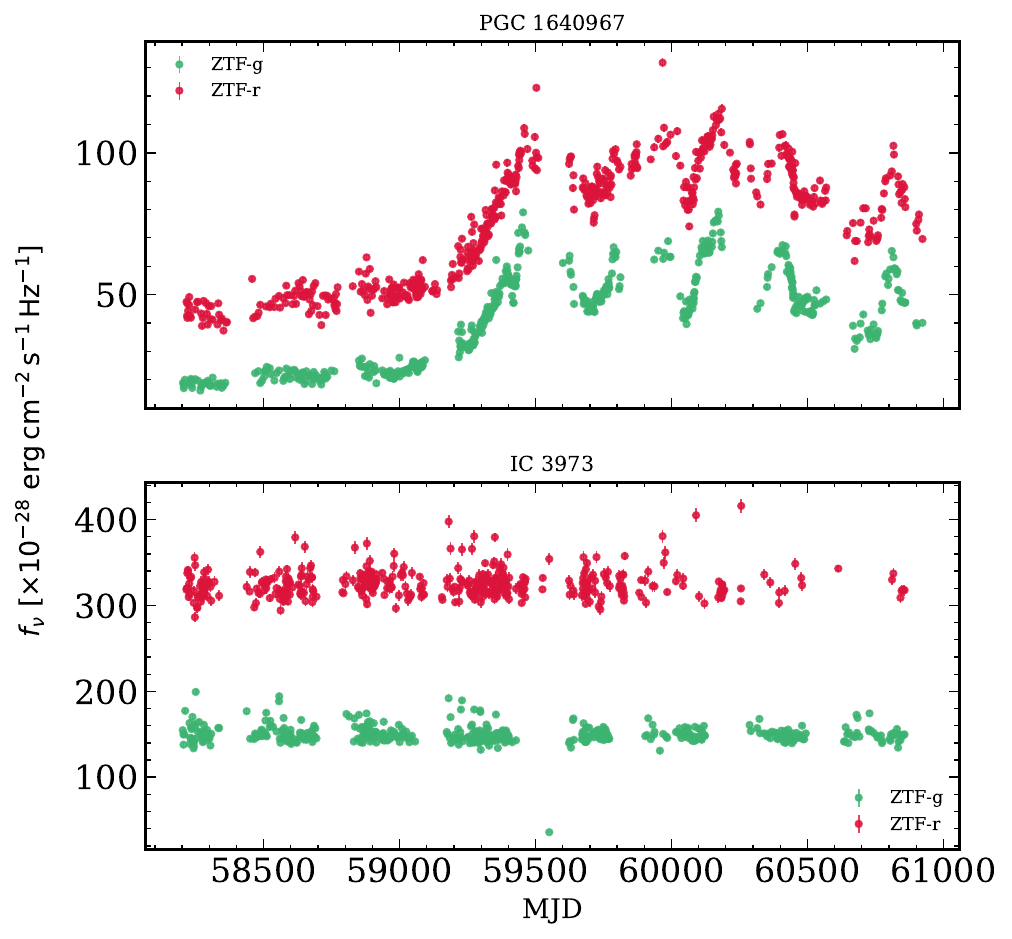}
    \caption{Representative cases in the $g_{2}$ vs K plane. The top panel shows an object with low Kurtosis and high K ($g_{2} \approx -1.04$, $K \approx 0.90$). Whereas the lower panel shows an object with high Kurtosis and low K ($g_{2} \approx 34.17$, $K \approx 0.45$). Provided values are measured on the g-band.}
    \label{fig:Kurtosis_K_extremes}
\end{figure}

Besides the scatter metrics, we also found a strong positive correlation of $\rho_{s} \sim 0.9$ between BRP and the kurtosis. This correlation is shown in the right panel (third from the top) of Figure \ref{fig:corrmatrix}. The observed relation is non-linear and asymptotic, which is a consequence of the BRP definition that limits its upper value to 1. This trend is produced because both metrics will show high values for distributions in which the fluxes are highly concentrated around the median, with a reduced number of outliers in the tails of the distribution. The significant impact of individual flux measurements in the Kurtosis is also reflected in the observed anti-correlation with the Stetson index K (bottom right panel in Figure \ref{fig:corrmatrix}). Although both metrics are intended to quantify the importance of the tails of the flux distribution, individual outliers have a much reduced impact on the value of K. This difference in the kurtosis estimators is illustrated in Figure \ref{fig:Kurtosis_K_extremes}, which compares the light curves of an object with high Kurtosis and modest K, with one of high K and modest Kurtosis. The object with high K (top-panel) shows high-amplitude variability throughout the observed period. In contrast, the high Kurtosis case is likely driven by a single data point close to $MJD\sim59500$. This data point might be an outlier, since we do not see a counterpart in the r-band. Visual inspection of additional cases in the Kurtosis vs K plane reveals the same behavior. This result highlights that the Stetson K index represents a more reliable metric to identify extreme variability events, whereas the extreme Kurtosis values should be interpreted with caution. \\

One of the strongest correlations in the set is observed between J and RoMS. Both metrics are constructed from the flux deviations normalized by the errors, but J additionally includes temporal information by measuring the coherence between subsequent fluctuations. The observed tight correlation suggests that this temporal information does not substantially modify the order predicted by RoMS. This might indicate that objects with large flux deviations also show a high degree of temporal coherence, leading to high RoMS and J values. Nevertheless, the observed dispersion increases towards large RoMS values, suggesting that sources with large variability amplitudes show a wider variety of temporal behaviors.

\subsection{Type I vs Type II variability}\label{subsec:type1_type2}
\begin{figure}
\centering
\includegraphics[width=\columnwidth]{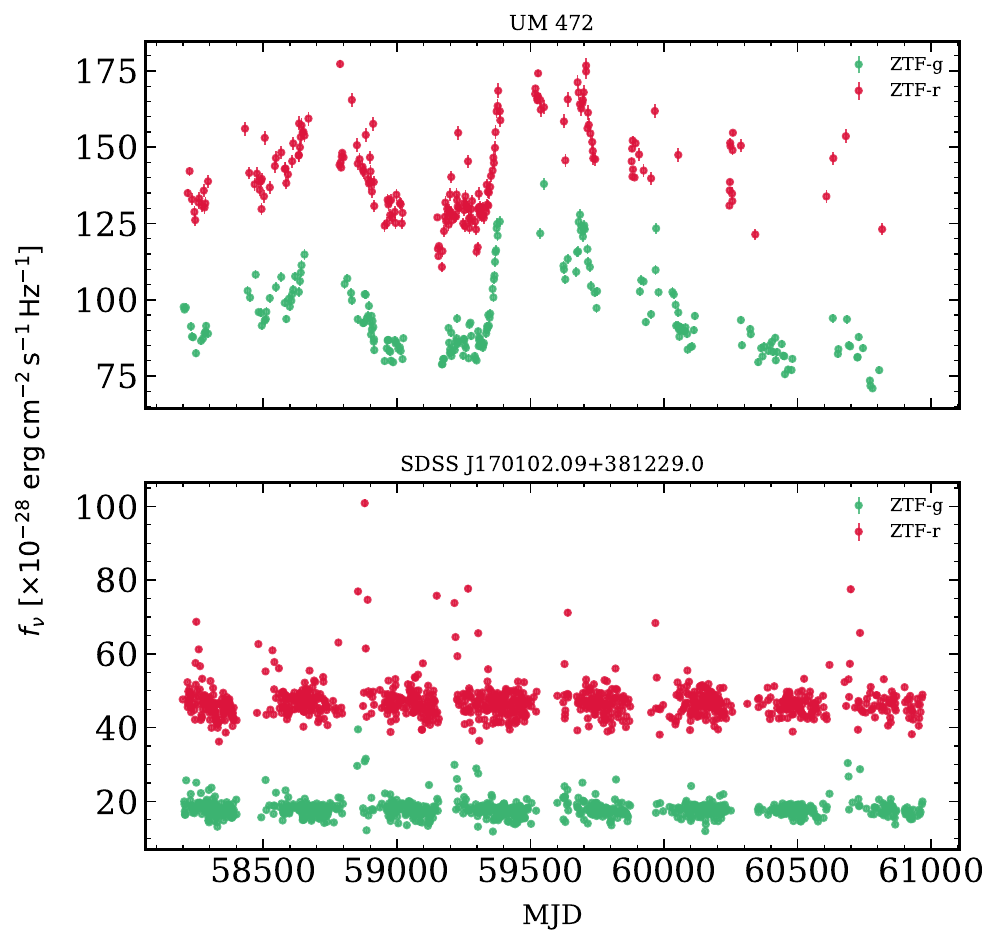}
    \caption{Example of two objects with very similar $F_{\rm var}$ values, but distinct variability behaviors. The object in the upper panel corresponds to a Type I AGN, with $F_{\rm var}\approx 0.0998$, while the object in the bottom panel corresponds to a Type II, with $F_{\rm var} \approx  0.0997$. Values correspond to the g-band.}
    \label{fig:Fvar_comparison}
\end{figure}

\begin{figure}
\centering
\includegraphics[width=\columnwidth]{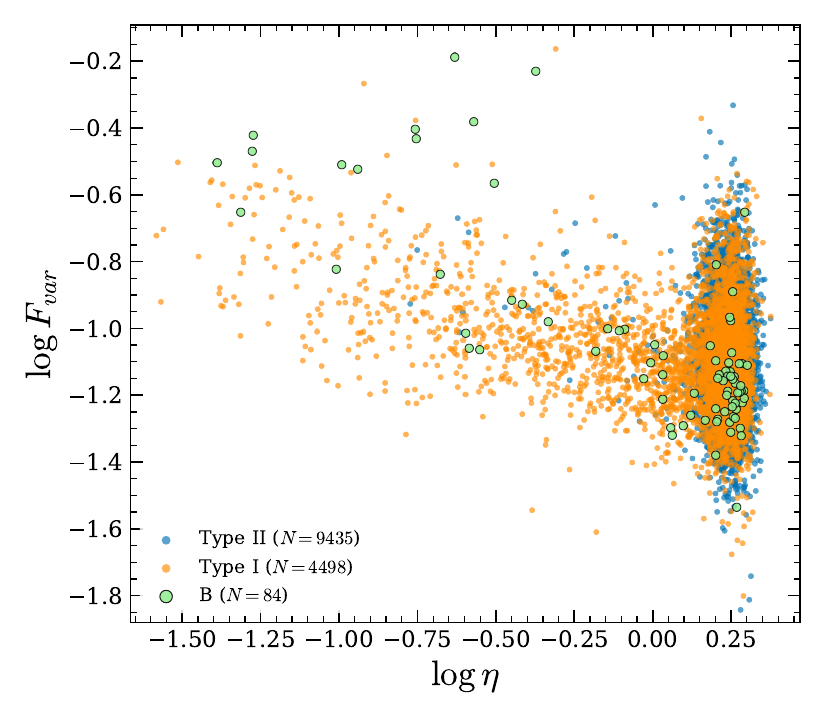}
    \caption{$F_{\rm var}$ vs $\eta$ plane. Objects are color-coded by their classification. Blue for Type II AGNs, orange for Type I, and green for BL Lac objects. The number of sources for each class is shown in the legend.}
    \label{fig:fvar_eta}
\end{figure}

\begin{figure}
\centering
\includegraphics[width=\columnwidth]{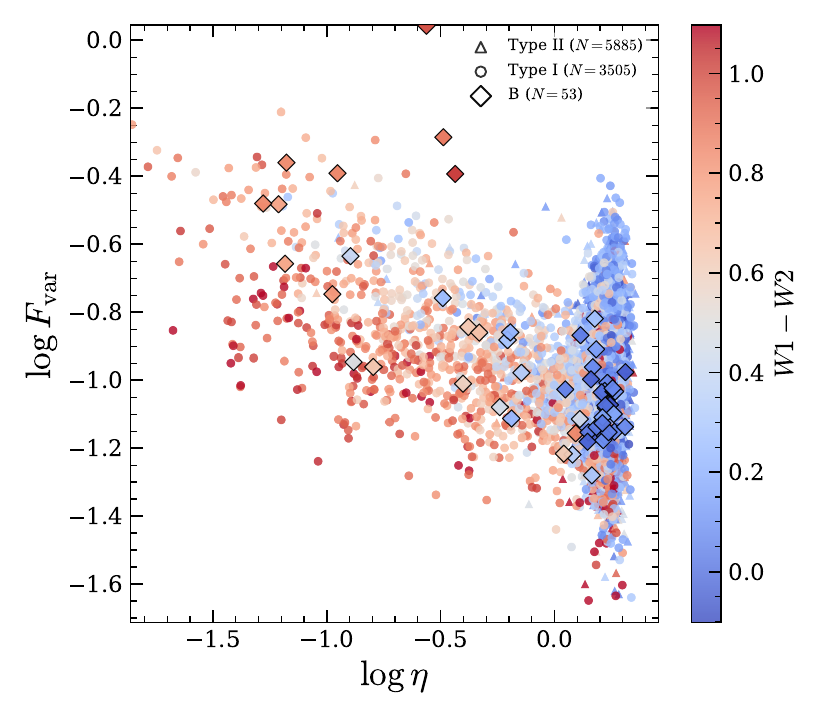}
    \caption{$F_{\rm var}$ vs $\eta$ plane colored using the WISE W1-W2 color of each object. The different AGN types are plotted with different markers, as shown in the legend. The variation in the number of sources with respect to Fig. \ref{fig:fvar_eta} is produced because not all objects had available WISE photometry.}
    \label{fig:fvar_eta_colors}
\end{figure}

We analyzed the differences between different AGN types in our metrics using the classification provided by Milliquas. \footnote{Milliquas originally provides the labels: \textbf{"A"}: AGN, \textbf{"Q"}: QSO, \textbf{"K"}: NLQSO, \textbf{"N"}: NLAGN, \textbf{"B"}: BL Lac, where "NL" means narrow-line Type II sources. We refer to "A" and "Q" as Type I, "K" and "N" as Type II, and "B" as Blazars}. 
As can be noticed in the scatter plots of Figure \ref{fig:corrmatrix}, there is no clear separation between Type I and Type II AGNs in the displayed correlations. In Type II AGNs, the obscuration of the central sub-parsec region reduces the AGN contribution. Therefore, we would expect these sources to show weaker signs of variability. However, most of the metrics in our set do not detect this difference. In particular, we found that the scatter metrics cannot distinguish between Type I AGNs with clear signs of variability and Type II AGNs with light curves affected by a few outliers; an example of this is presented in Figure \ref{fig:Fvar_comparison}. This indicates that the identification of AGN-like variability using scatter metrics alone may not be accurate. Instead, we have found that the inclusion of the $\eta$ metric represents a better strategy to uncover AGN-like variability, as we present in Figure \ref{fig:fvar_eta}.
The $F_{\rm var}-\eta$ plane shows a clear bimodality, with a region of overlap between Type I, Type II, and blazars at high $\eta$ values, and a tail at low $\eta$ populated almost exclusively by Type I and blazar sources. We have found that the objects in the tail show AGN-like variability, with significant fluctuations that are coherent in time, following the damped random walk behavior. However, the sources in the overlap region show unclear signs of variability. A visual inspection of Type I AGNs in the overlap region reveals that they do not show clear signs of variability. Instead, their light curves do not differ from those of Type II AGNs. It is possible that the AGN contribution in these objects is weaker and affected by a variable degree of obscuration of the central region. The variability features in these obscured cases may be diluted by the host-galaxy emission. This interpretation is supported by Figure \ref{fig:fvar_eta_colors}, where we have color-coded each object in the $F_{\rm var}-\eta$ plane using the WISE W1-W2 color. This color is known to correlate with the AGN fraction of a galaxy \citep{hickox2017}. Specifically, a higher W1-W2 color often indicates a larger contribution from AGN-heated dust relative to host-galaxy starlight, suggesting stronger nuclear emission. We found that the overlap region is populated by objects with low W1-W2, indicating that the AGN contribution is weak. The color value gradually increases towards lower $\eta$ values, supporting the interpretation that the tail is populated by AGN-dominated objects. The significant fraction of Type I objects in the overlap region may suggest that some of these sources could be misclassified.\\

We noticed that the $F_{\rm var}-\eta$ plane we present in Figure \ref{fig:fvar_eta}, resembles the $\sigma_{QSO}-\sigma_{var}$ of \citet{wang2024}. These metrics were originally introduced by \citet{butler2011} as a way to identify quasar-like variability. $\sigma_{\rm var}$ quantifies whether a source shows variability, while $\sigma_{QSO}$ assess whether the observed variability follows a DRW behavior \citep{wang2024}. In our plane, $F_{\rm var}$ works similarly to $\sigma_{var}$, but we emphasize that $F_{\rm var}$ can be affected by outliers, as we show in Figure \ref{fig:Fvar_comparison}, which may explain why the overlap region in our plane extends to high values on $F_{\rm var}$. On the other hand, our plane suggests that $\eta$ is the metric that discerns whether the variability is consistent with a DRW. $\eta$ is defined as the ratio between the variation of subsequent flux measurements and the global dispersion in the light curve. In a damped random walk process, the fluctuations have a temporal memory given by the damping timescale $\tau$. Therefore, flux measurements separated in time by $\Delta t << \tau$ are expected to be similar, reducing the variation between subsequent flux measurements, leading to lower values of $\eta$. The rest-frame damping timescale of optical AGN variability is typically tens to hundreds of days, with a characteristic value of $\tau_{\rm rest}\sim200\,\mathrm{d}$ reported for SDSS Stripe~82 quasars \citep{macleod2010,Burke2021}. Longer-baseline quasar monitoring has yielded median values of several hundred days \citep{Stone2022}. The corresponding observer-frame timescales, $\tau_{\rm obs}=(1+z)\tau_{\rm rest}$, are therefore much longer than the typical ZTF cadence of 1-3 days in our data. Thus, DRW variability will naturally produce low $\eta$ values in our sample. However, we emphasize that other variability behaviors, such as a monotonic decrease/increase in flux, as observed in CL events, will also produce very low values of $\eta$. This is the case of the presented object in the upper panel of Figure \ref{fig:outlier_eta}.

\subsection{Uniform Manifold Approximation and
Projection Analysis} \label{subsec:umap}

\begin{figure*}[t]
\centering
\includegraphics[width=\textwidth]{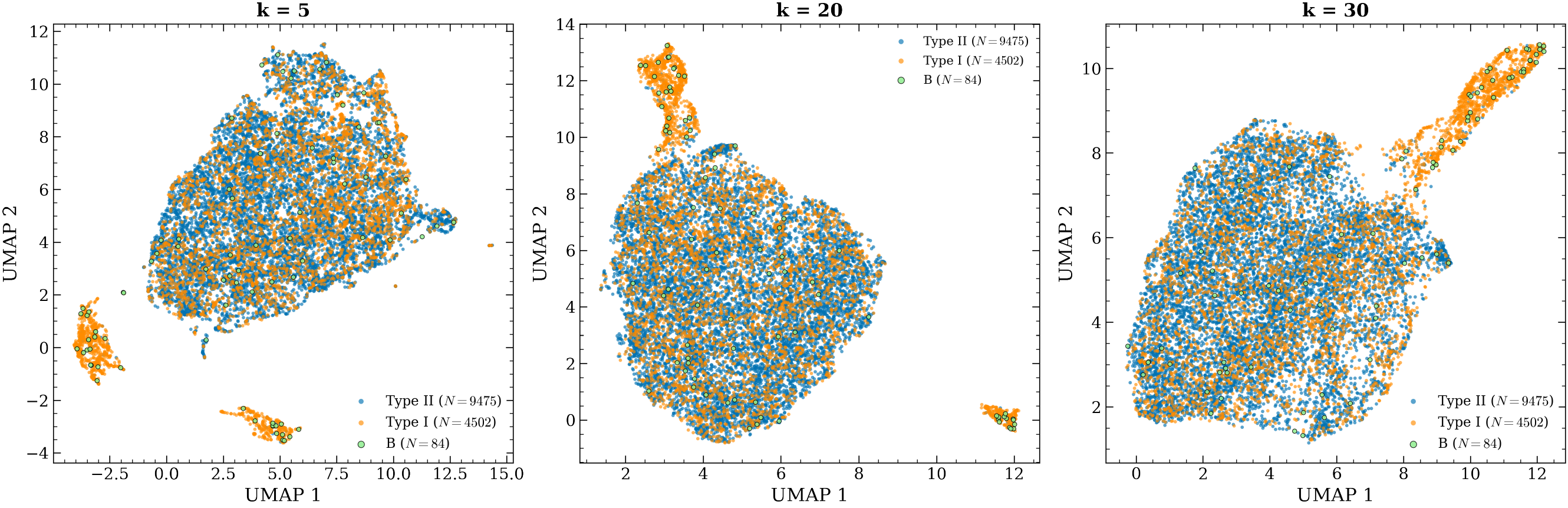}
    \caption{UMAP representation of our data for different values of k. Objects are color-coded by their classification. Blue for Type II AGNs, orange for Type I, and green for BL Lac objects. The number of sources for each class is shown in the legend. The number of neighbors k is shown in the title of each panel.}
    \label{fig:umap_class}
\end{figure*}

\begin{figure}[t]
    \centering
\includegraphics[width=\columnwidth]{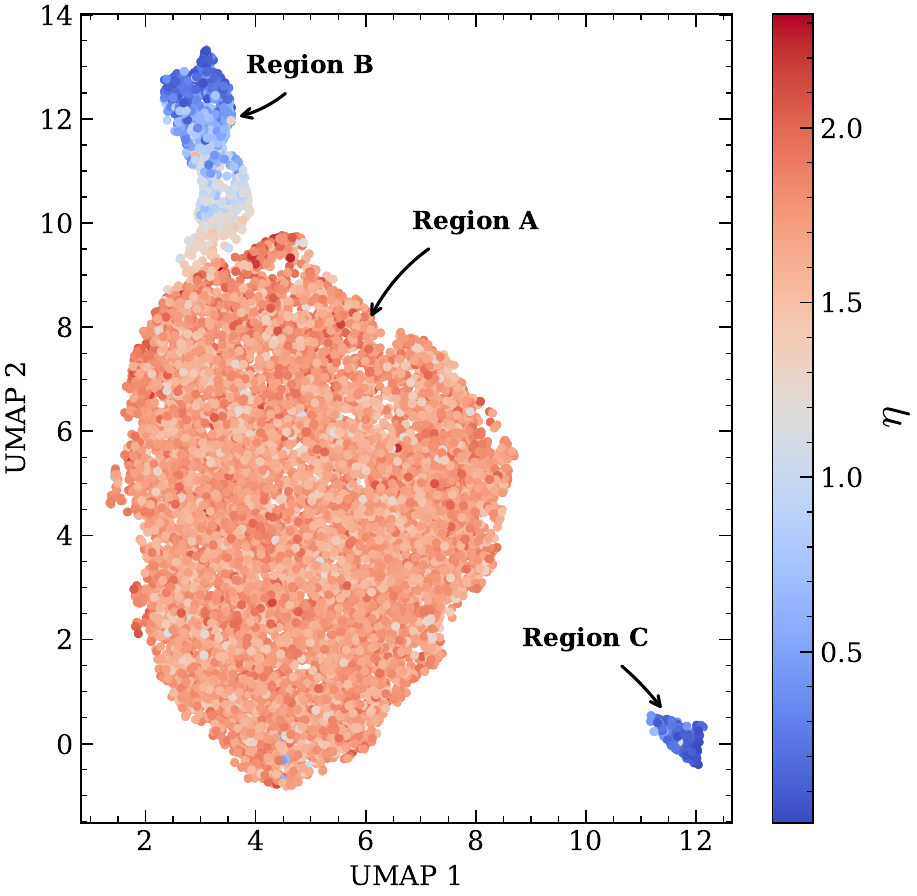}
    \caption{2D representation of the UMAP colored using the $\eta$ metric. We highlight the position of the three main regions.}
    \label{fig:umap_eta}
\end{figure}

We explored the realization of a Uniform Manifold Approximation and Projection \citep[UMAP,][]{mcinnes2018} with our set of metrics. The UMAP algorithm initially builds a graph representation of the geometric structure formed by the data points in the high-dimensional space defined by the metrics. This graph is constructed by finding the k-nearest neighbors to each data point, and its connections are weighted by the strength of the local relationships between points. Therefore, this graph provides a fuzzy topological representation of the original data, which encodes the similarity relations between objects. The algorithm then maps the relations of this graph into a low-dimensional representation that preserves the neighborhood structure of the original space, facilitating the visualization of clusters of objects with similar variability behaviors. \\

We used the UMAP implementation of the Python package ${\tt umap}$ \citep{mcinnes2018umap-software}. For the calculations, we used the following subset of metrics: $\eta$, MAD, RoMS, $g_{1}$, Mid80, K, BRP, $R_{cs}$, Linear, $\Delta BB$, and $\chi$. The metrics in this subset were selected to capture distinct aspects of the variability, reducing any redundancy in the input data. The analysis was done using metrics calculated on the g-band. \\

We compressed our set of metrics into two dimensions, UMAP1 and UMAP2. The primary UMAP hyperparameters are the distance metric, the minimum distance, and k (n\_neighbors). In this work, we adopted the standard Euclidean distance metric, the default minimum distance of 0.1, and random initialization with a random seed of 0. Regarding $k$, we explored the values $k=5,15,30$. The $k$ parameter regulates the number of nearest neighbors that are considered to make the connections in the high-dimensional graph. A lower value means that we are connecting each data point only to its closest neighbors. This approach increases the sensitivity to highly local relations between objects, but can artificially produce disconnected regions of objects, whose variability does not significantly differ from the rest. In contrast, a higher value means that each object is connected to more neighbors, including objects that may be farther. This approach is better for revealing global structures in the dataset, but can miss substructures with physical meaning.\\

We present the results of our UMAP analysis in Figure \ref{fig:umap_class}.
For all $k$ values, we identified a central cluster where Type I and Type II objects show significant overlap, along with smaller regions dominated by Type I objects. Blazars appear similarly distributed across the different regions. This result is similar to the structures we found in the 
$F_{\rm var}-\eta$ plane (see Fig.  \ref{fig:fvar_eta}). \\

In the particular case of $k$=30, we observe a single region dominated by Type I AGNs, which seems to be an extension of the bigger central region, rather than a disconnected structure. In contrast, the map for $k$=20 shows two regions dominated by Type I AGNs. The upper left one is connected to the bigger central region, while the lower right one is isolated. Similarly, the map for $k$=5 produces two disconnected structures dominated by Type I AGNs. The results shows that smaller values of $k$ allows to separate the single structure we observed in $k$=30, into two smaller regions, which likely present different behaviors. In the following, we focused on the analysis of the UMAP for $k$=20.\\

We identified the three main regions in the $k$=20 UMAP as Region A (center), Region B (upper left), and Region C (lower right). To better interpret the variability properties in these regions, we color-coded the UMAP using the $\eta$ metric, as shown in Figure \ref{fig:umap_eta}.
Objects in Region A display almost exclusively $\eta$ values greater than 1.0. Therefore, they lie in the overlap region of the $F_{\rm var}-\eta$ plane. In comparison, Region B shows a color gradient towards lower values as we move farther from Region A. Region C shows the lowest $\eta$ values, suggesting smooth, temporally correlated fluctuations. Objects in B and C likely correspond to the Type I tail we observed in Fig. \ref{fig:fvar_eta}. The observed color gradient in Region B suggests that this structure is an extension of Region A, while Region C seems to be an independent structure.\\

\begin{figure*}
    \centering
    \includegraphics[width=0.9\textwidth]{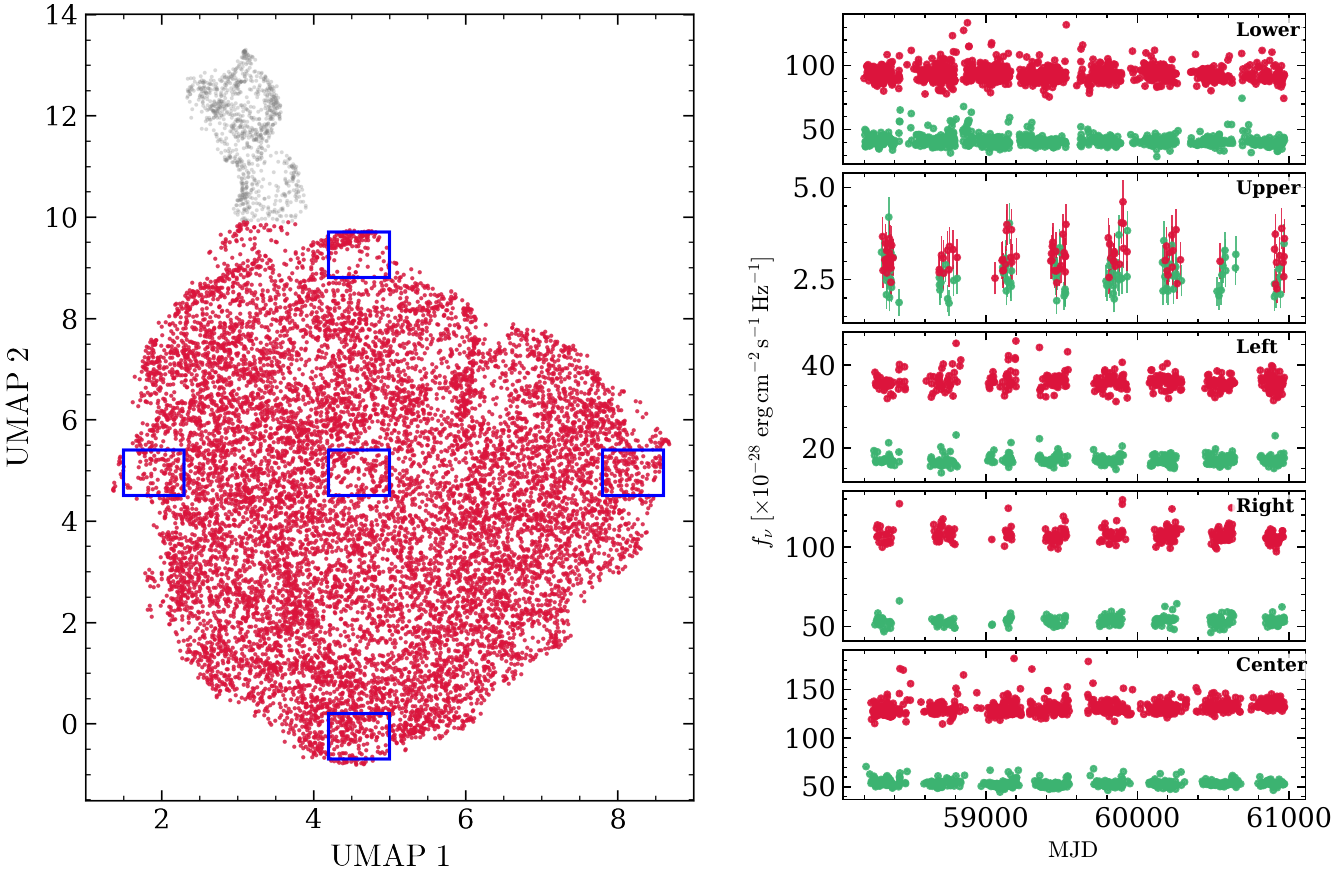}
    \caption{Representative light curves in Region A. Left: approximated location of Region A in red color, along with the remaining structures in grey color. We divided Region A into 5 parts highlighted with blue rectangles, from which we extracted the representative light curves. Right: Representative light curves for each rectangle in Region A; we labeled them according to the position of the rectangle on the left panel.}
    \label{fig:umap_lc_A}
\end{figure*}

\begin{figure*}
    \centering
    \includegraphics[width=0.9\textwidth]{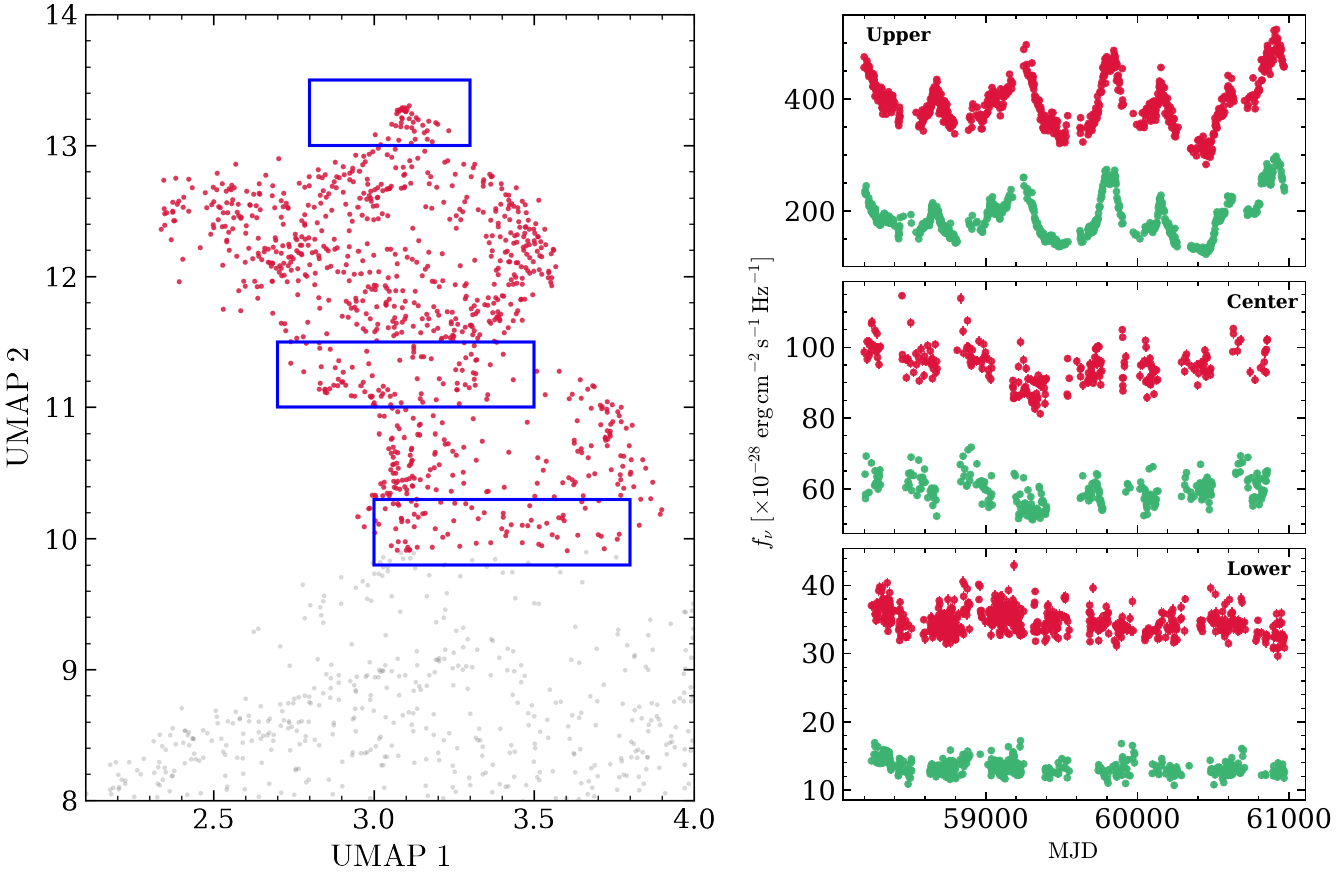}
    \caption{Same as \ref{fig:umap_lc_A} for Region B.}
    \label{fig:umap_lc_B}
\end{figure*}

\begin{figure*}
    \centering
    \includegraphics[width=0.9\textwidth]{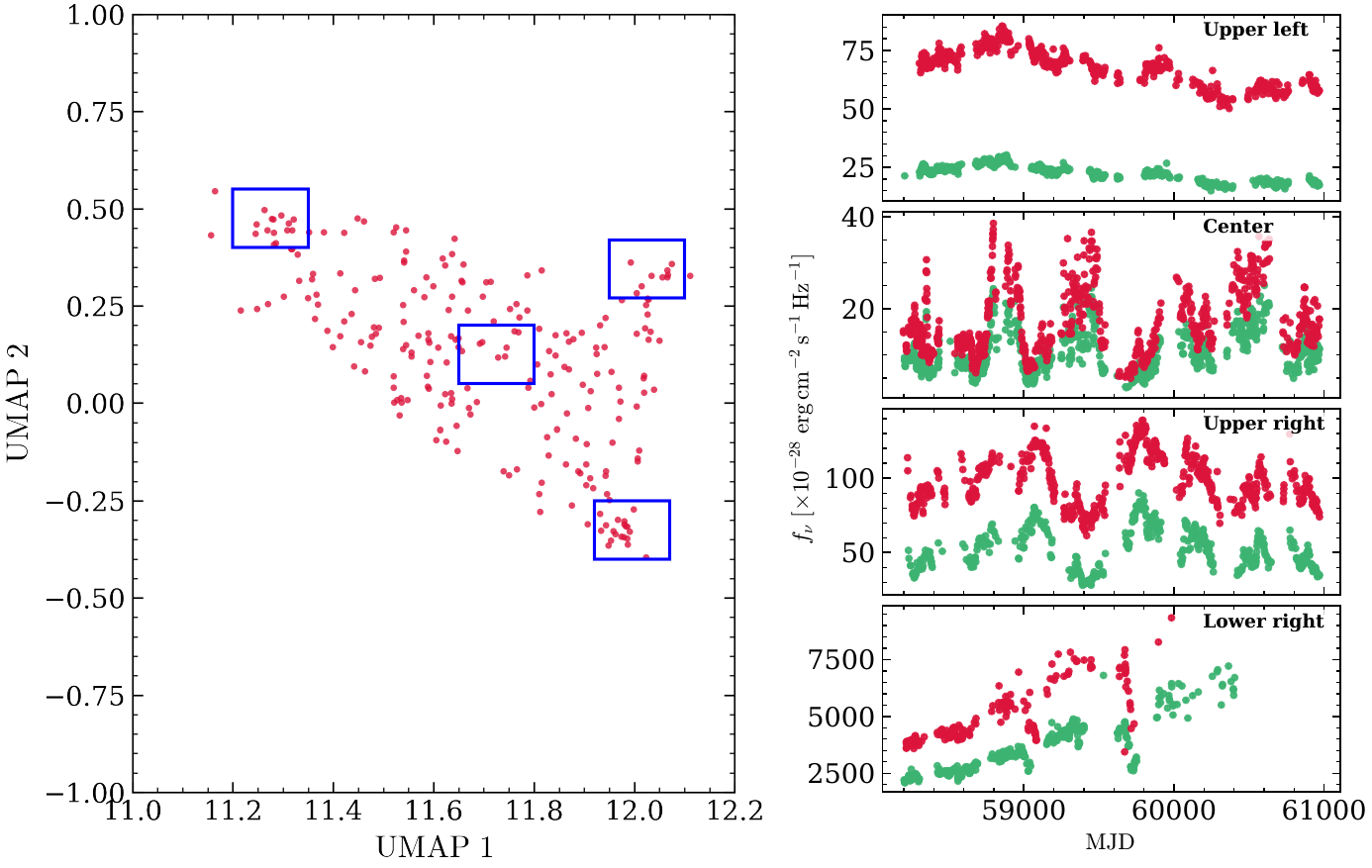}
    \caption{Same as \ref{fig:umap_lc_A} for Region C.}
    \label{fig:umap_lc_C}
\end{figure*}

We visually inspected the light curves of the objects occupying each region, and we present representative cases in Figures \ref{fig:umap_lc_A}, \ref{fig:umap_lc_B}, and \ref{fig:umap_lc_C}.
We found that objects in Region A preferentially show weak signs of variability and sparse light curves, which explains the observed high $\eta$ values. We did not find significant variations between objects across this region, but we detected that objects in the upper part, close to Region B, tend to show larger photometric errors. Our selected metrics are not intended to measure the quality of the errors, but these can have an effect on some of them. This is the particular case of RoMS, as its definition makes use of the photometric errors (see Eq. \ref{equation:ROMS}).\\

\begin{figure}
    \centering
    \includegraphics[width=\columnwidth]{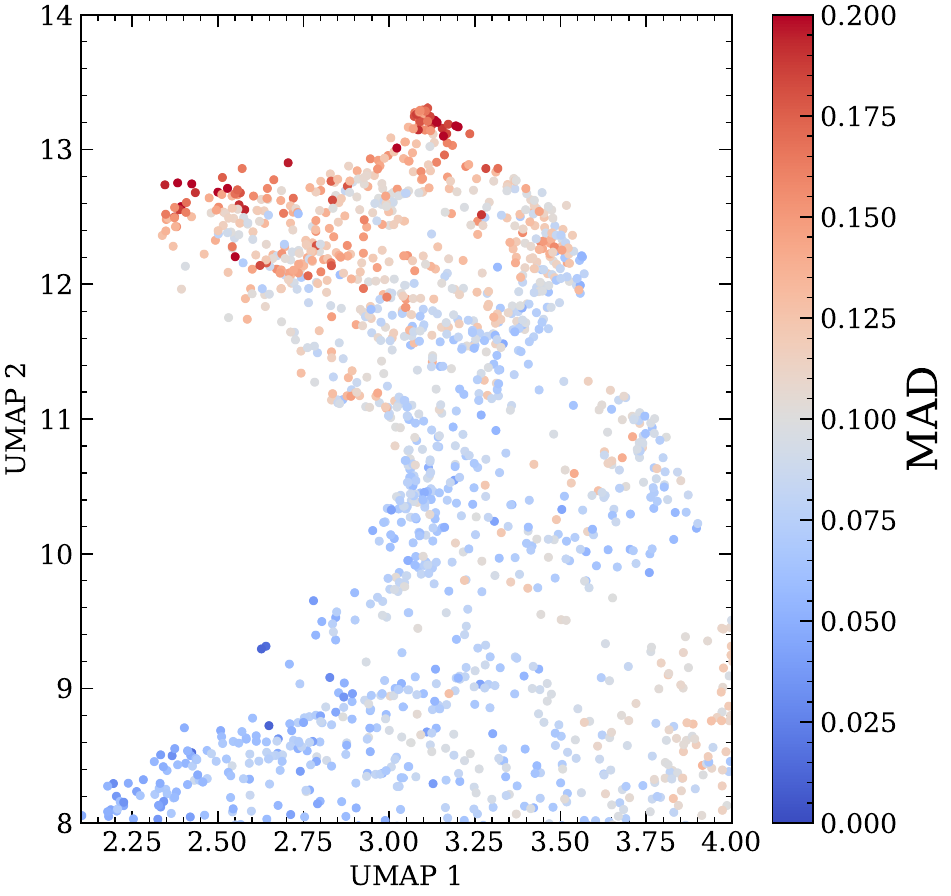}
    \caption{Region B colored using the MAD metric.}
    \label{fig:umap_B_MAD}
\end{figure}

Figure \ref{fig:umap_lc_B} shows the representative objects in Region B. We detect that objects in the lower part of this region tend to show similar light curves to those observed in Region A, with weak variability amplitudes. The signs of variability gradually increase as we move farther from Region A, with objects in the upper part showing high-amplitude, coherent fluctuations. This evolution in the variability amplitude is confirmed by the MAD metric, which shows a gradient towards higher values in the upper part of Region B, as shown in Figure \ref{fig:umap_B_MAD}.\\
Figure \ref{fig:umap_lc_C} shows the representative cases of Region C. In this case, we found a higher diversity of behaviors. Objects in the upper left part of the region tend to show long-term variability trends, which may resemble the behavior of the non-variable cases in Region A. Objects in the center and upper right parts of the region show high variability amplitudes, while objects in the lower right position additionally display monotonic trends. We highlight that the changing-look object NGC 4151 (Fig. \ref{fig:outlier_linear}) lies in this lower right corner (see also the bottom right panel in Figure \ref{fig:umap_lc_C}).\\ 

\begin{figure*}
    \centering
    \includegraphics[width=0.9\textwidth]{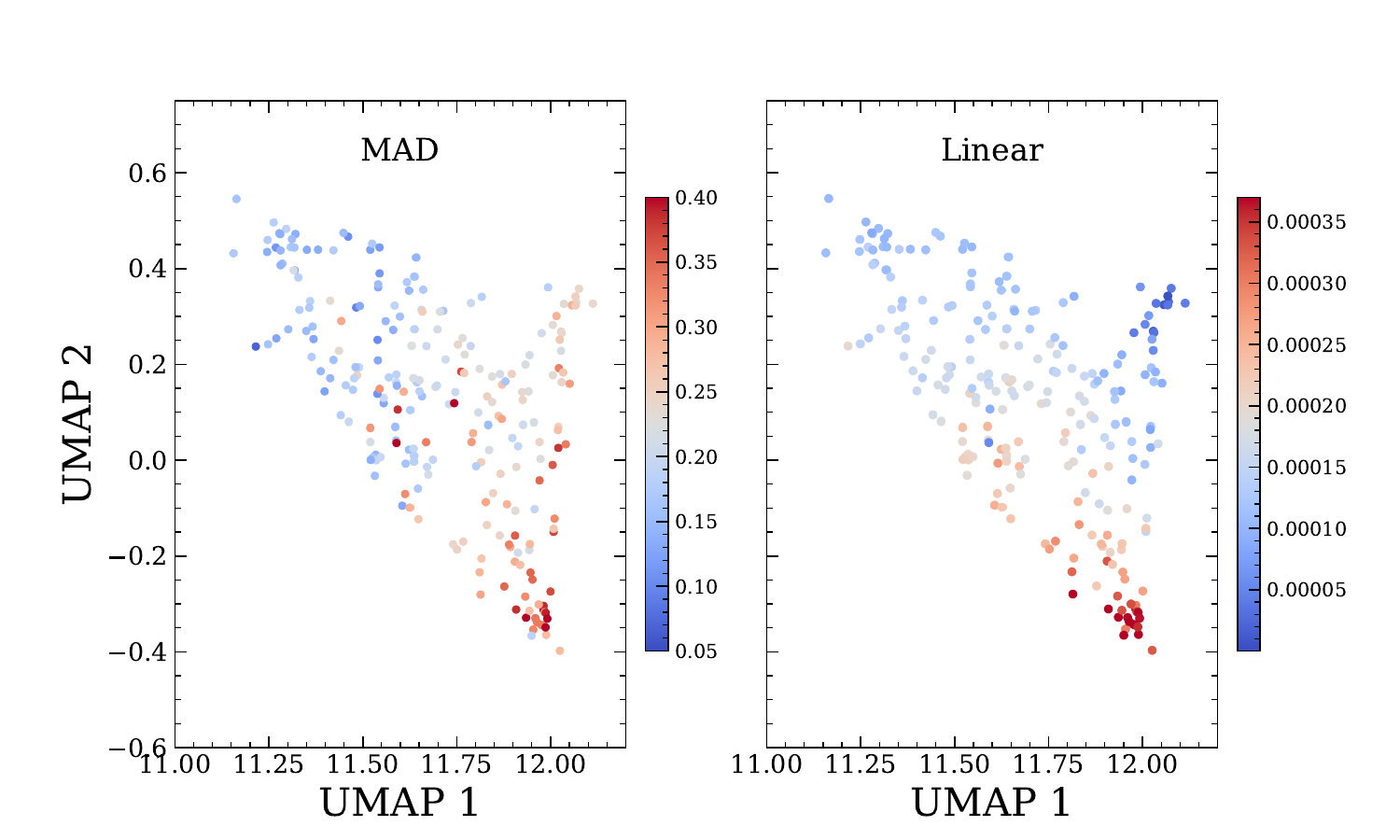}
    \caption{Left: Objects in Region C colored with the MAD metric. Right: Same as left using the Linear metric.}
    \label{fig:regionC_mad_linear}
\end{figure*}

Objects in Region C display the highest MAD values in the sample, and a subset of them also show high values in the Linear metric (see Fig. \ref{fig:regionC_mad_linear}), suggesting monotonic changes in brightness. 
This behavior resembles that of the Extreme Variable Quasar population \citep[EVQs,][]{rumbaugh2018}, which displays high-amplitude variability, with a subset of them showing CL behavior. To test whether objects in Region C can be classified as EVQs, we followed \citet{Macleod2016} criteria and labeled all objects that satisfy the criterion $\Delta g > 1.0$ as EVQs, where $\Delta g = |g_{2}-g_{1}|$, and $g_{1}$ and $g_{2}$ are any two epochs in the light curve. We found that objects in Region C show preferentially high $\Delta g$ values. In particular, the region shows an ensemble value of $\langle \Delta g \rangle \sim 1.11 $, and 55\% of the objects satisfy the EVQ criterion, compared to only 7\% and 18\% for Regions B and A, respectively. This result highlights that Region C is preferentially populated by extreme variability events. Even though we detected a fraction of EVQs in Regions B and A, the number is likely contaminated by false EVQs, as the light curves in these regions show weak signs of variability and are significantly affected by single outlier points (see the examples in Figure \ref{fig:umap_lc_A}).    


\section{Discussion}\label{sec:discussion}
While the metric approach offers a simple and bulk-applicable technique to characterize AGN light curves, its interpretation and connection with the fundamental AGN properties should be treated with caution, as multiple effects can impact the final value for a given metric. Our metric values were obtained for an AGN sample from Milliquas under specific redshift and baseline conditions using only light curves from ZTF DR24. In the following, we assess the potential impact of these conditions on our reported values, aiming to provide the necessary context to compare our catalog of values with future works using different samples, redshift ranges, and photometric campaigns.

\subsection{Baseline effect}
AGNs are intrinsically variable sources. Therefore, our metrics are time-dependent parameters. The presented values in our catalog are not fixed quantities, and they were exclusively derived using the 8-year light curves from ZTF DR24. The values are expected to vary if a different baseline is considered. We test the stability of our metrics against baseline variations. To do this, we created shorter-baseline versions of our light curves by trimming the light curves at 1, 2, 4, and 6 years. We recomputed the metrics for the shorter-baseline versions, and we compared them with the value obtained using the complete light curve (8 years). To quantify the variation for the complete sample, we used the coefficient of determination $R^{2}$ 
\begin{equation}
    R^{2} = 1- \frac{\sum_{i} (y_{i}-\hat{y}_{i})^{2}}{\sum_{i} (y_{i}-\bar{y}_{i})^{2}}
\end{equation}
where $y_{i}$ corresponds to the value of the metric for each object using the complete baseline (8 years) and $\hat{y_{i}}$ is the predicted value of the metric using the shorter baseline versions. An $R^{2}$ close to 1 indicates that the predictions $\hat{y_{i}}$ are in good agreement with the reference values. 
\begin{figure*}
    \centering
    \includegraphics[width=0.9\textwidth]{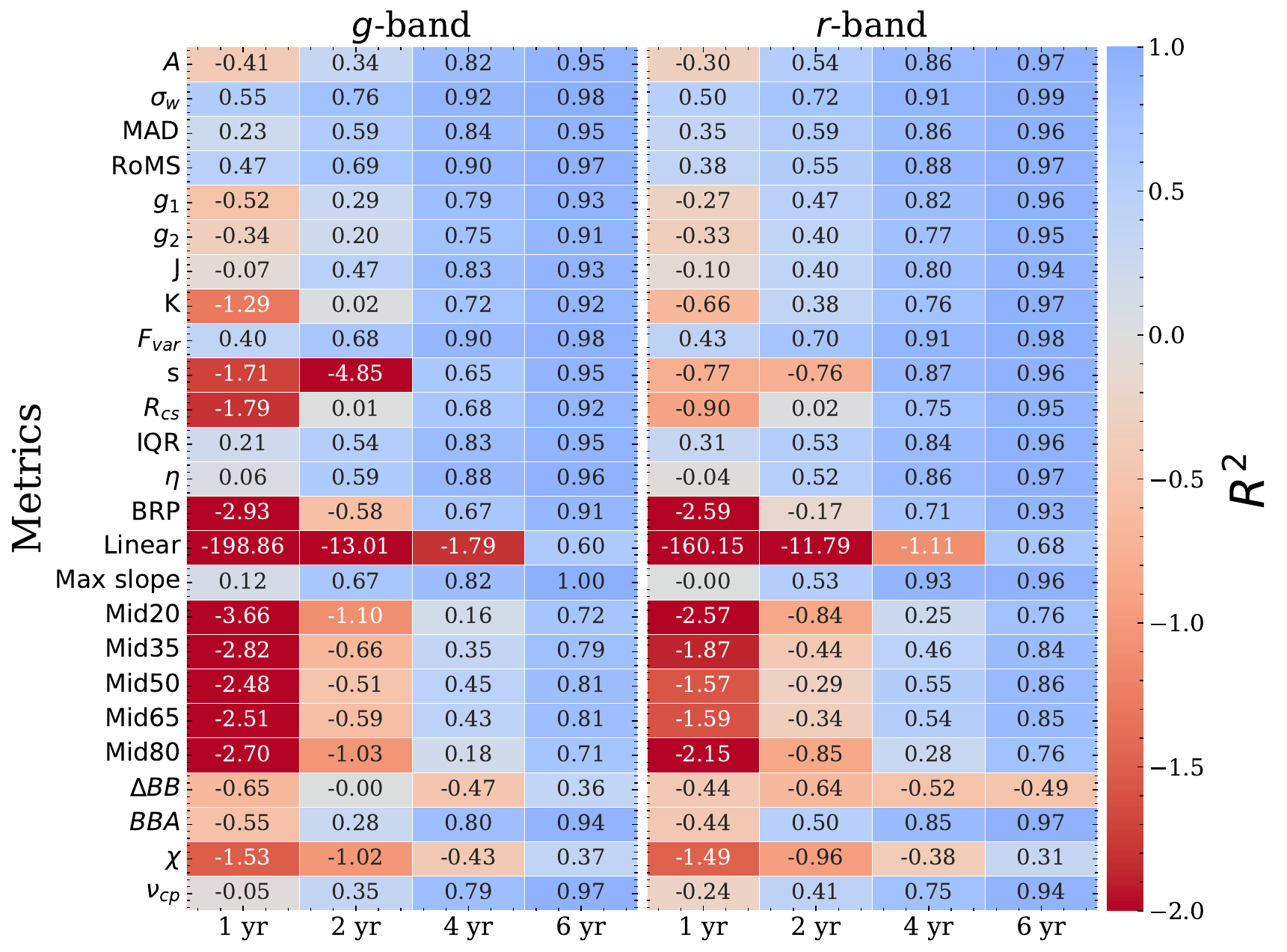}
    \caption{Baseline effect on variability metrics. The heatmap shows the variation in the metric value for each studied baseline. Matrix elements are colored using the $R^{2}$ value. The left panel shows the resulting heatmap for the g-band light curves, while the right panel shows the result for the r-band.}
    \label{fig:baseline_eff}
\end{figure*}\\

The results are shown in the heatmap of Fig. \ref{fig:baseline_eff}. We can immediately notice that some metrics are more stable than others. Most scatter metrics, such as $F_{\rm var}$ or MAD, are able to recover the original metric value using half of the baseline, i.e., 4 years. Metrics that measure flux distribution properties, like the percentile ratios or $K$, tend to require a longer baseline of 6 years to obtain a similar accuracy. Significant variations are observed in the $\mathrm{Linear}$ metric, which shows an extreme sensitivity to baseline. The metric is only able to recover the reference value after covering 6 years of observations. In a similar way, $\Delta BB$ and $\chi$ show strong variations with baseline, and in their particular case, the shorter baseline light curves are not able to recover the reference value, with the $R^{2}$ value being less than 0.5 even after 6 years. We suggest that these metrics are mostly driven by particular events in the light curve, and their value may not have a relation with the intrinsic AGN properties. On the contrary, the stability of scatter metrics like $F_{\rm var}$ suggests that they might have a stronger connection with the fundamental AGN properties. For example, it has been found that the $F_{\rm var}$ metric correlates with the AGN luminosity and Eddington ratio \citep{benati2025}.\\

Overall, we found that for most metrics, the obtained value using the 6-year light curves is similar to that obtained using the full light curve (8 years). Therefore, it is possible that the presented values in our catalog will not show significant variations within 2 years. We emphasize that this result is similarly observed in the g and r bands; despite minor variations, the overall conclusion remains similar. 

\subsection{Studied redshift range}\label{subsec:redshift_range}
The presented analyses were performed using nearby AGNs ($z \lesssim 0.09$). This redshift threshold was imposed to limit differences in rest-frame wavelength coverage and sampling timescales across the ZTF light curves. Our studied light curves reflect the variations in the optical range around the $H\beta$ and $H\alpha$ regions. Whether our metrics will show different values if a different wavelength regime is considered is something that needs to be studied further. In addition, the photometry of nearby AGNs can be affected by a higher contribution of the host galaxy in the flux measurements. The contribution of the host can have the effect of diluting the AGN variability features and modifying the values of some metrics, as we have explored in our previous work \citep{Martinez2026}.
\begin{table}
\centering
\caption{Metrics values before and after host subtraction for Mrk 493. Variations are calculated using the QSO+Host value as reference.}
\label{tab:host_effect}

\scriptsize
\setlength{\tabcolsep}{2.8pt}
\renewcommand{\arraystretch}{0.92}

\begin{tabular}{lcccccc}
\hline\hline
Metric
& \multicolumn{2}{c}{QSO+Host}
& \multicolumn{2}{c}{Only QSO}
& \multicolumn{2}{c}{$\Delta$ (\%)} \\
\cline{2-3} \cline{4-5} \cline{6-7}
& $g$ & $r$ & $g$ & $r$ & $g$ & $r$ \\
\hline
$A$                   & 0.649 & 0.511 & 0.841 & 0.795 & +29.6 & +55.5 \\
$\sigma_w$            & 0.125 & 0.097 & 0.162 & 0.150 & +29.6 & +55.5 \\
$\mathrm{MAD}$        & 0.131 & 0.099 & 0.170 & 0.153 & +29.6 & +55.5 \\
$\mathrm{RoMS}$       & 9.266 & 7.060 & 9.266 & 7.060 & +0.0  & +0.0  \\
$g_1$                 & 0.629 & 0.579 & 0.629 & 0.579 & -0.0  & +0.0  \\
$g_2$                 & -0.209 & -0.172 & -0.209 & -0.172 & +0.0 & -0.0  \\
$\mathrm{J}$          & 1.756 & 1.494 & 1.756 & 1.494 & -0.0  & +0.0  \\
$\mathrm{K}$          & 0.808 & 0.797 & 0.808 & 0.797 & -0.0  & +0.0  \\
$F_{\mathrm{var}}$    & 0.137 & 0.104 & 0.178 & 0.161 & +29.6 & +55.5 \\
$\mathrm{s}$          & 3.69e-4 & 5.02e-4 & 6.19e-4 & 0.001 & +67.9 & +141.9 \\
$R_{\mathrm{cs}}$     & 0.367 & 0.321 & 0.367 & 0.321 & +0.0  & -0.0  \\
$\mathrm{IQR}$        & 0.181 & 0.135 & 0.234 & 0.210 & +29.6 & +55.5 \\
$\eta$                & 0.086 & 0.136 & 0.086 & 0.136 & +0.0  & +0.0  \\
$\mathrm{BRP}$        & 0.402 & 0.434 & 0.402 & 0.434 & +0.0  & +0.0  \\
$\mathrm{Linear}$     & 1.64e-5 & 9.11e-6 & 2.13e-5 & 1.42e-5 & +29.6 & +55.5 \\
$\mathrm{Max\ slope}$ & 0.088 & 0.092 & 0.115 & 0.144 & +29.6 & +55.5 \\
$\mathrm{Mid20}$      & 0.134 & 0.123 & 0.134 & 0.123 & +0.0  & +0.0  \\
$\mathrm{Mid35}$      & 0.266 & 0.233 & 0.266 & 0.233 & -0.0  & -0.0  \\
$\mathrm{Mid50}$      & 0.402 & 0.387 & 0.402 & 0.387 & -0.0  & +0.0  \\
$\mathrm{Mid65}$      & 0.585 & 0.565 & 0.585 & 0.565 & +0.0  & +0.0  \\
$\mathrm{Mid80}$      & 0.776 & 0.778 & 0.776 & 0.778 & -0.0  & +0.0  \\
$\Delta BB$           & 0.017 & 0.029 & 0.023 & 0.045 & +29.6 & +55.5 \\
$\mathrm{BBA}$        & 0.582 & 0.454 & 0.754 & 0.706 & +29.6 & +55.5 \\
$\chi$                & 0.000 & 0.000 & 0.000 & 0.000 & --    & --    \\
$\nu_{cp}$            & 0.039 & 0.037 & 0.039 & 0.037 & +0.0  & +0.0  \\
\hline\hline
\end{tabular}
\end{table}
We analyzed the impact of the host galaxy on our metrics using Mrk 493 as a case study. This nearby AGN (z = 0.031) represents a suitable case where both AGN and host contributions are significant. We retrieved the DESI DR1 optical spectra of this object \citep{desi2025}, derived the host contribution with {\tt PyQSOfit} \citep{pyqsofit}, and convolved the derived host spectra with the ZTF filters using {\tt SNCosmo} \citep{Barbary2016}, as better described in \citet{Martinez2026}. The resulting variations in each metric after host subtraction are shown in Table \ref{tab:host_effect}. \\

We found that most metrics are insensitive to the host contribution. This is the case for flux distribution metrics and most temporal correlation metrics, such as K, $R_{cs}$, or $\eta$. Several metrics in the set show exactly the same variation. This is a trivial effect caused by the normalization applied to the light curves before the metric calculation. To illustrate this, we can consider the case of the Amplitude metric. This metric is actually not affected by the host subtraction, since the host contribution is just a constant that gets canceled in the definition:
\begin{align*}
    A_{qso+host} &= f_{max}-f_{min} \\
    &= (f_{max}^{qso} + F_{host}) - (f_{min}^{qso} + F_{host}) \\
    &= f_{max}^{qso}-f_{min}^{qso}
\end{align*}
However, since the calculation of A is done using the normalized light curve, the actual value of A is given by:
\begin{align*}
    A = \frac{f_{max}^{qso}-f_{min}^{qso}}{\overline{F}}  
\end{align*}
where $\overline{F}$ is the mean flux of the light curve. Because the host contribution is constant, the value of $\overline{F}$ in the original light curve is $\overline{F}=\overline{F_{qso}} + F_{host}$, while for the host-subtracted light curve it is $\overline{F} = \overline{F_{qso}}$. Thus, the fractional variation in A reduces to:
\begin{align*}
    \Delta \% &= \frac{A_{qso} - A_{qso+host}}{A_{qso+host}} \times 100 \\
    &= \frac{\overline{F_{qso+host}}}{\overline{F_{qso}}}-1 \times 100 \\
    &= \frac{F_{host}}{\overline{F_{qso}}} \times 100
\end{align*}
For Mrk 493, the ratio $\frac{F_{host}}{\overline{F_{qso}}} \times 100$ takes a value of 29.59\% and 55.54\% for the g and r bands, respectively. Therefore, the metrics that show these variations are not actually affected by the host, but they vary due to the change in the normalization factor. The only metric that deviates from this behavior is the smoothness parameter $s$. Unlike the other metrics, the variations in $s$ have a quadratic dependence on the ratio $\frac{F_{host}}{\overline{F_{qso}}}$ (see Eq. \ref{eq:smooth}). In this case, the fractional variation scales as:
\begin{equation}
    \Delta \% = \left[\left( 1 + \frac{F_{host}}{\overline{F_{qso}}}\right)^{2}-1\right]
\end{equation}
For Mrk 493, this results in fractional variations of 67.94\% and 141.93\% in the g and r bands, respectively, matching the obtained values.\\

We found that most of our metrics are not affected by the host galaxy emission, and the cause of the observed variations is the previous normalization of the light curves. We have shown that the effect of this normalization can be trivially derived from the metric definitions, and therefore easily quantified.           

\subsection{Comparison with ZVAR DR1 and QZO}
The ZTF Variability Archive (ZVAR) \citep{graham2026ztf} recently provided metric measurements for around 1.3 billion variable sources, including some metrics that are also present in our pipeline. These measurements are calculated on ZTF forced photometry light curves covering observations until July, 2023. In the following, we compare the ZVAR measurements with those provided in this work.\\

We retrieved ZVAR measurements for AGNs from the QZO catalog \citep{nakoneczny2025}. QZO consists of around 5 million quasars observed with ZTF. Unlike our Milliquas samples, the QZO catalog is expected to be dominated by bright Type I AGNs, allowing us to study the impact of the sample properties on the metrics. \\

We queried the ZVAR API for 4,849,574 quasar coordinates from QZO using a radius of $1''$. We found 4,199,794 sources with available ZVAR measurements. Each object can have multiple ZTF light curves in ZVAR; for our analysis, we only kept the light curves with the highest number of photometric measurements in the g and r bands, and we additionally required a minimum of 100 exposures in each band. We used the available QZO photometric redshifts to further restrict the sample to nearby AGNs $(z_{phot} < 0.09)$, matching the limit we used in our catalog (Section \ref{sec:data}). This constraint significantly reduces the number of objects to only $\sim$ 990 sources. Given the significant uncertainty of photometric redshift measurements, we decided to relax the constraint and explore a second sample with $(z_{phot} < 0.9)$. This higher redshift sample consists of 221,089 objects. \\

\begin{figure*}[t]
    \centering
    \includegraphics[width=0.9\textwidth]{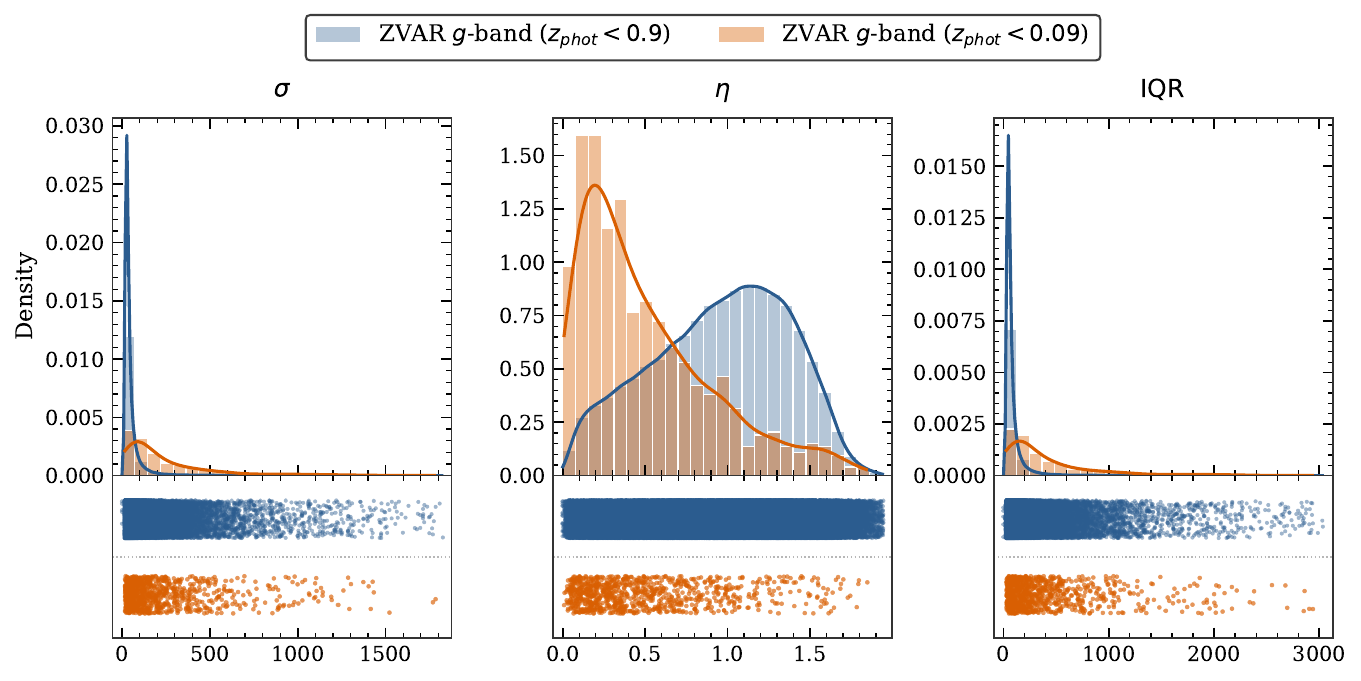}
    \caption{Metric distribution for the ZVAR samples. The top of each panel shows the probability density histograms (bars) and the kernel density estimate (KDE, solid curve). The bottom shows the actual metric values for each source dispersed along the x-axis with a jitter for visualization.}
    \label{fig:zvar_dist}
\end{figure*}

We analyzed the ZVAR measurements for the $\eta$, IQR, and $\sigma$ metrics. The corresponding distributions are presented in Fig. \ref{fig:zvar_dist}.  
The distributions of the scatter metrics $\sigma$ and IQR show pronounced tails; this effect is particularly extreme for the $(z_{phot} < 0.9)$ sample, which shows kurtosis values of $g_{2} \sim 20000$ in both distributions. The $(z_{phot} < 0.09)$ sample is less extreme, showing kurtosis values of $g_{2} \sim 200$. Nevertheless, the kurtosis values for both samples are significantly higher than those observed in our sample, $g_{2}\sim 7.7$ and $g_{2}\sim 31$ for $\sigma_{w}$ and IQR, respectively.\\

The distributions for $\eta$ show clear differences with respect to our sample, Fig. \ref{fig:metric_dist}. The $(z_{phot} < 0.09)$ sample is dominated by objects with $\eta < 1.0$, contrary to our distribution, which is dominated by objects at $\eta > 1.0$. The $(z_{phot} < 0.9)$ sample shows more similarity to our sample; its distribution peaks around $\eta \sim 1.0-1.5$, but it also contains a high fraction of objects at $\eta < 1.0$.\\ 

The higher fraction of objects with low $\eta$ values suggests that the QZO samples are dominated by objects with strong AGN activity, as previously anticipated. This could also explain the pronounced tails in the distributions of the scatter metrics, which can be produced by high-amplitude variability events. The significant differences with our sample highlight the impact of the sample properties in the metric distributions. \\

\begin{figure*}[t]
\centering
\includegraphics[width=\textwidth]{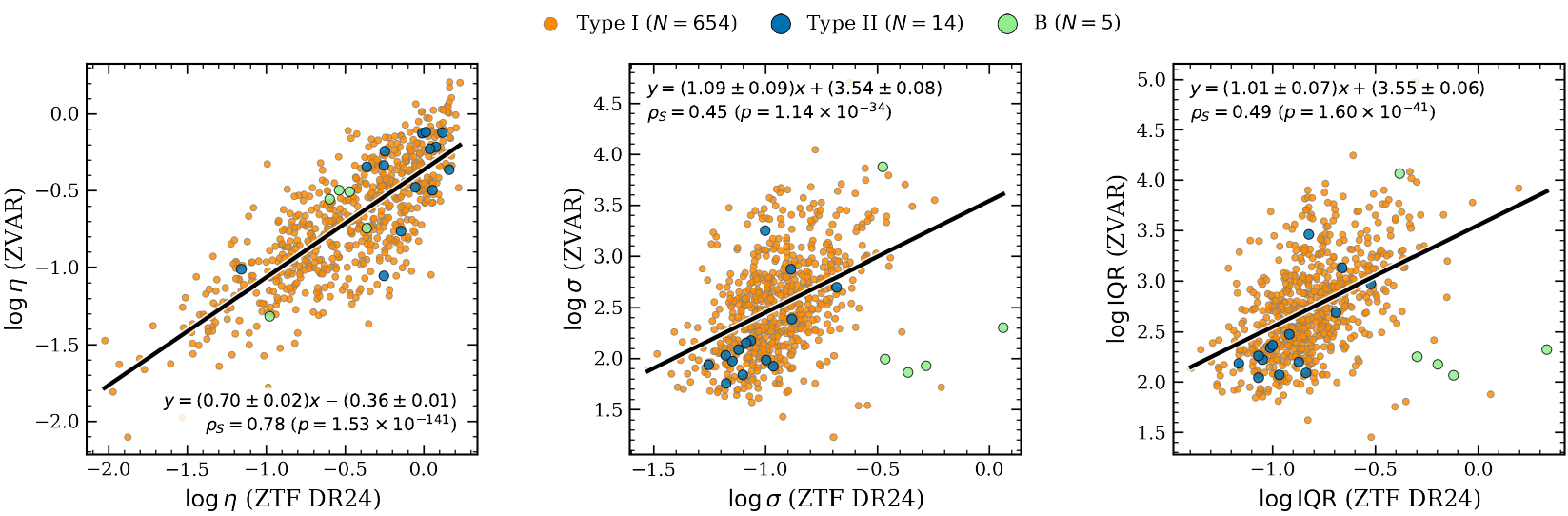}
    \caption{Comparison between the variability metrics estimated with our pipeline using ZTF DR24 light curves (x-axis) and those from ZVAR using forced photometry (y-axis). Objects are color-coded by their classification. Blue for Type II AGNs, orange for Type I, and green for BL Lac objects. The number of sources for each class is shown in the legend. A linear fit to the data is shown with a black solid line. The fitted linear equation, Spearman correlation coefficients, and p-values are shown inside each panel.}
    \label{fig:zvar_lin}
\end{figure*}

We also compared the exact ZVAR values with those provided in our catalog. For this, we cross-matched the $(z_{phot} < 0.9)$ sample with our catalog using a radius of 1.0'', finding 673 matches. Since the ZVAR light curves include observations only up to July, 2023, we restricted our ZTF DR24 light curves to the same time interval and recalculated the metrics. The comparison is presented in Fig. \ref{fig:zvar_lin}.
For the comparison, we computed the Spearman correlation coefficients and p-values with {\tt scipy.stats} module \citep{2020SciPy-NMeth}, and fitted a linear model to the data using the Ordinary Least Squares (OLS) method implemented in the {\tt statsmodels} library \citep{seabold2010}. We found good agreement between the $\eta$ measurements, with a strong correlation coefficient of $\rho_{s} \sim 0.8$ and a slope of $\sim 0.7$ in the linear fit. The different AGN classes similarly distribute around the best linear relation. On the other hand, the scatter metrics $\sigma$ and IQR show more modest correlations, and we noticed that the few Blazars in the sample seem to deviate more from the linear relation. As opposed to $\eta$, the $\sigma$ and IQR metrics depend on the flux scale of the objects, and since the ZVAR metrics are computed on difference fluxes from forced photometry, we expected differences in the scale of the ZVAR values compared to our catalog. We confirmed these differences and suggest that they are potentially responsible for the observed significant scatter. 

\section{Summary and Looking ahead} \label{sec:summary}
We have developed a Python-based metric variability pipeline compiling multiple algorithms from the literature. We have tested its use to characterize 14,061 nearby AGNs ($z \lesssim 0.09$) from the Milliquas catalog \citep{Flesch2023}. The main findings of this analysis can be summarized as follows: 
\begin{itemize}

    \item Most amplitude and scatter metrics are strongly correlated. For instance, $F_{\rm var}$ and $\sigma_w$ exhibit a Spearman correlation coefficient of $\rho_s \approx 0.9$. The analysis of outliers in their distributions revealed potential major flare events; see Figures \ref{fig:outlier_fvar} and \ref{fig:outlier_MAD}. These findings highlight the potential of metrics to identify interesting science cases for follow-up. 
    
    \item Some amplitude/scatter metrics can give similar values for very different light curves; this is the case for ($F_{\rm var}$), which shows similar values for flare-like events and cases that are likely affected by a single deviant epoch, Fig.\ref{fig:Fvar_comparison}. Including temporal metrics like $\eta$ helps to distinguish these behaviors, providing a way to filter out unreliable events.

    \item Adopting the classification provided by Milliquas, we studied the differences in our metrics between Type I, Type II, and Blazars. We found that these classes significantly overlap in most studied correlations, Fig. \ref{fig:corrmatrix}. We highlight the particular case of the $(F_{\rm var})-(\eta)$ plane, Fig. \ref{fig:fvar_eta}. This plane is characterized by two main regions: at $\eta \lesssim 1.0$ we only observe Type I and Blazars, while at $\eta \gtrsim 1.0$ all the classes overlap. The undetected differences in the variability of different AGN types in the overlap region invite a review of the classifications of these objects, as they could have changed during the ZTF monitoring. We highlight the necessity to keep AGN classes updated to ensure an interpretation of the metric trends. Future repeated spectroscopic observations with 4MOST \citep{swann2019,frohmaier2025} or DESI \citep{desi2022} will be crucial for this task.

    \item We interpreted the two behaviors in the $(F_{\rm var})-(\eta)$ plane by including infrared photometry from WISE, Fig. \ref{fig:fvar_eta_colors}. We found a clear gradient where objects towards low $\eta$ values show redder W1-W2 colors. The WISE W1-W2 is known to correlate with the AGN fraction of a galaxy \citep{hickox2017}. Therefore, we suggest that the two observed behaviors in the plane reflect differences in the strength of the AGN activity.  
    
    \item ¿Can we identify different AGN behaviors from photometry alone using metrics? Motivated by this idea, we explored a higher-dimensional analysis using UMAP. This non-linear dimensionality reduction technique is well-suited to capture the complex relations that we observed between our metrics. Our main result is that the UMAP embedding identifies a region enriched in sources with large-amplitude variability, which contains a sub-region with objects showing long-term flux trends, as shown in Figure \ref{fig:umap_lc_C}. A significant fraction of these objects satisfy the criterion of extremely variable quasars (EVQs) \citep{rumbaugh2018}. Further study of this region can help to prioritize candidates for spectroscopic follow-up to test for new changing-look transitions.

    \item Inspired by our previous work, \citet{Martinez2026}, we assessed the impact of the observed baseline in our metrics. A baseline truncation analysis reveals that, on a population level, amplitude and scatter metrics show good agreement with their eight-year reference values after about four years of monitoring. Flux distribution metrics like the percentile ratios or the Stetson (K) generally require around six years to show a similar result, and Linear and some Bayesian-block metrics show higher sensitivity to baseline; this is summarized in Fig. \ref{fig:baseline_eff}. We conclude that shorter baselines can still have predictive power for some metrics, and therefore our catalog of values can be used as a reference for future studies using extended baselines.
    
    \item We provide the metric analysis results as a value-added catalog, described in Table \ref{tab:catalog_description_1}. We will make our pipeline code publicly available as soon as the publication is accepted.
    
\end{itemize}

Our method opens several possibilities for future analysis. The comparison of our results against an extended baseline with previous photometric surveys such as CRTS \citep{drake2009} or ASAS-SN \citep{Shapee2014}, is a natural next step. We also intend to explore the relation between metrics and the fundamental SMBH properties in the future. For this, large spectroscopic catalogs, such as SDSS DR16 \citep{wu2022} or, more recently, DESI DR1 \citep{wu2026}, are suitable complements. Looking forward, the Vera C. Rubin Observatory Legacy Survey of Space and Time \citep[LSST,][]{ivezic2019} represents the next revolution for the study of AGN variability. LSST will survey the southern sky in six filters (ugrizy) covering wavelengths from the near-ultraviolet to the near-infrared, providing an unprecedented volume of multi-band, multi-epoch photometric data (see \citet{Panda2026} for a detailed review). We highlight that the presented metric pipeline is dynamic and survey-agnostic and can be applied directly to LSST light curves. \\

\begin{acknowledgments}
This research was part of the NOIRLab Chile REU program funded by NSF Award Number 2349023. DMC acknowledges funding from ANID, Chile, through the scholarship ANID MAGISTER NACIONAL, 22251176, and financial support from Millenium Nucleus TITANs (NCN2023${\_}$002, AIM23-0001), and the China-Chile Joint Fund (CCJRF2310). SP is supported by the International Gemini Observatory, a program of NSF NOIRLab, which is managed by the Association of Universities for Research in Astronomy (AURA) under a cooperative agreement with the U.S. National Science Foundation, on behalf of the Gemini partnership of Argentina, Brazil, Canada, Chile, the Republic of Korea, and the United States of America. 

\end{acknowledgments}

\facilities{PO:48, WISE, Mayall}


\software{numpy \citep{numpy}, matplotlib \citep{matplotlib}, astropy \citep{2013A&A...558A..33A,2018AJ....156..123A},
pandas \citep{2010scpy.soft.....M},
scipy \citep{2020SciPy-NMeth},
umap \citep{mcinnes2018umap-software},
statsmodels \citep{seabold2010},
PyQSOFit \citep{pyqsofit},
SNCosmo \citep{Barbary2016}}

\bibliography{main}{}
\bibliographystyle{aasjournalv7}

\appendix

\renewcommand{\thetable}{\thesection.\arabic{table}}
\setcounter{table}{0}

\renewcommand{\thefigure}{\thesection.\arabic{figure}}
\setcounter{figure}{0}

\section{Variability Metrics}
\label{app:metrics}
\subsection{Weighted Mean flux ($\mu_{w}$)}
Mean of the flux distribution weighted by the flux errors
\begin{equation}
    \bar{F} = \frac{\sum_{i} w_{i}F_{i}}{\sum_{i}w_{i}} 
    \label{equation:mean}
\end{equation}
Where $w_{i}$ are the weights obtained using the flux errors $\sigma_{i}$:
\begin{equation}
    w_{i} = \frac{1}{\sigma_{i}^{2}}
    \label{equation:weights}
\end{equation}
\subsection{Amplitude (A)}
The amplitude of the variability for a given light curve is defined as the difference between the highest and lowest flux states:
\begin{equation}
    A = F_{max}-F_{min}
    \label{equation:amp}
\end{equation}.
This metric has been used for the selection of AGN candidates \citep{garcia2015}. 
\subsection{Weighted Standard deviation ($\sigma_{w}$)}
Standard deviation of the flux distribution weighted by the flux errors, as defined in \citet{sokolovsky2017}.
\begin{equation}
    \sigma_{w} = \sqrt{\frac{\sum_{i} w_{i}}{\left(\sum_{i} w_{i}\right)^{2}- \sum \left(w_{i}^{2}\right)} \sum_{i} w_{i}\left(F_{i}-\bar{F}\right)^{2}}
\label{equation:std}
\end{equation}
\subsection{Median absolute deviation (MAD)}
The median absolute deviation \citep{hampel1974} is another metric to quantify the amplitude of the variability and has the advantage of being robust against outliers. It's defined as the median of the absolute differences between the flux and the median flux:
\begin{equation}
    MAD = median (|F_{i}-F_{median}|)
    \label{equation:MAD}
\end{equation}

\subsection{Robust median statistic (RoMS)}
The RoMS statistic \citep{enoch2003} quantifies the deviation of the fluxes with respect to the median, normalized by the noise. This parameter has been used in the past to detect significant variable objects \citep{rose2007,Burdanov2014}. 
\begin{equation}
    ROMS = \frac{1}{n-1} \sum_{i} \frac{|F_{i}-median(F_{i})|}{\sigma_{i}}
    \label{equation:ROMS}
\end{equation}
The expected value of RoMS for a non-variable source is 1.
\subsection{Mean variance}
The mean variance metric is the simple ratio between the standard deviation and the mean flux $\frac{\sigma}{\bar{F}}$. This metric is expected to be large in objects with significant variability amplitudes.

\subsection{$F_{var}$}
The $F_{var}$ metric \citep{vaughan2003} provides a robust way to quantify the intrinsic variability by taking into account the observational noise.
\begin{equation}
    F_{var} = \sqrt{\frac{\sigma_{XS}^{2}}{ \bar{F}^{2}}}
    \label{equation:FVAR}
\end{equation}
Where $\sigma_{XS}$ corresponds to the excess variance, defined as:
\begin{equation}
    \sigma_{XS}^{2} = S^{2}-\bar{\sigma}^{2} 
\end{equation}
with 
\begin{equation}
    S^{2} = \frac{1}{n-1} \sum_{i} \left(F_{i}-\bar{F}\right)^{2}
\end{equation}
This metric has been extensively used in AGN variability studies, such as \citet{Edelson2002,Peterson2004}, and more recently in \citet{benati2025}.
\subsection{Interquartile range (IQR)}
The interquartile range \citep{kim2014} represents a robust metric to characterize the strength of the variability. In a similar way to the MAD, this metric is robust against outliers in the flux distribution
\begin{equation}
    IQR = Q_{3}-Q_{1}
\end{equation}
Where $Q_{1}$ and $Q_{3}$ are the first and third quartiles of the flux distribution, respectively.
\subsection{Skewness ($g_{1}$)}
The Skewness is the third standardized moment of a distribution. It measures the asymmetry of the flux distribution with respect to the mean value. A positive value means that the flux distribution is dominated by high flux states with respect to the mean, while a negative value points towards low flux states dominating the variability.
\begin{equation}
    g_{1} = \frac{\frac{1}{N} \sum_{i}\left(F_{i}-\bar{F}\right)^{3} }{\left[\frac{1}{N}\sum_{i} \left(F_{i}-\bar{F}\right)^{2}\right]^{3/2}}
\end{equation}
\subsection{Kurtosis ($g_{2}$)}
The Kurtosis is the fourth standardized moment of a distribution. It quantifies the importance of the tails in a flux distribution. A high positive value suggests the presence of extreme outliers:
\begin{equation}
    g_{2} = \frac{\frac{1}{N} \sum_{i}\left(F_{i}-\bar{F}\right)^{4} }{\left[\frac{1}{N}\sum_{i} \left(F_{i}-\bar{F}\right)^{2}\right]^{3/2}} - 3
\end{equation}
Notice that our definition corresponds to the excess Kurtosis. Under this convention, a Gaussian flux distribution will have a Kurtosis equal to zero. 
\subsection{Percentile ratios: (Mid20, Mid35, Mid50, Mid65, and Mid80)}
The percentile ratios were introduced by \citet{Richards2011} as a non-periodic feature to classify variable objects. These ratios characterize the shape of the flux distribution using its percentiles; they are defined as follows.

\begin{equation}
    Mid20 = \frac{P_{60}-P_{40}}{P_{95}-P_{5}}
\end{equation}
\begin{equation}
    Mid35 = \frac{P_{67.5}-P_{32.5}}{P_{95}-P_{5}}
\end{equation}
\begin{equation}
    Mid50 = \frac{P_{75}-P_{25}}{P_{95}-P_{5}}
\end{equation}
\begin{equation}
    Mid65 = \frac{P_{82.5}-P_{17.5}}{P_{95}-P_{5}}
\end{equation}
\begin{equation}
    Mid80 = \frac{P_{90}-P_{10}}{P_{95}-P_{5}}
\end{equation}
Where $P_{i}$ corresponds to the i-th flux percentile.
\subsection{Stetson J (J) }
The Stetson index J \citep{Welch1993} measures the degree of correlation between subsequent flux measurements, against what is expected from pure noise. A positive value is indicative of intrinsic real variability. The metric is defined as:
\begin{equation}
    J = \frac{\sum w_{i}\,sgn (\delta_{i}^{2}-1) \sqrt{|\delta_{i}^{2}-1|}}{\sum w_{i}}
    \label{eq:Jindex}
\end{equation}
where sgn returns the sign of the value, $\delta_{i}$ is defined as 
\begin{equation}
    \delta_{i} = \frac{f_{i}-\bar{f}}{\sigma_{f,i}}\sqrt{\frac{n}{n-1}}
    \label{eq:delta}
\end{equation}
and $w_{i}$ is:
\begin{equation}
    w_{i} = \left[1+\left(\frac{\delta_{i}}{2}\right)^{2}\right]^{-1}
    \label{eq:weight}
\end{equation}
In the equations, $f_{i}$ is the \textit{i}$^{\rm th}$ flux element, $\bar{f}$ the mean flux, $\sigma_{f,i}$ the \textit{i}$^{\rm th}$ uncertainty, and \textit{n} the number of data points. The Stetson J metric is commonly used to identify significant variability \citep{Ma2024,Martinez2026}.
\subsection{Stetson K (K)}
The Stetson index K \citep{stetson1996} was designed to be a robust estimator of the kurtosis for observational flux distribution. Quantifies the importance of tails in the flux distribution with respect to the average variability amplitude. This metric is defined as:
\begin{equation}
    K = \frac{\frac{1}{n}\sum_{i}|\delta_{i}|}{\sqrt{\frac{1}{n}\sum_{i}\delta_{i}^{2}}}
    \label{eq:kindex}
\end{equation}
Where $\delta_{i}$ was already defined in Equation \ref{eq:delta}. A Gaussian flux distribution would result in a K value of $\sim 0.798$. A higher value is indicative of extreme events in the flux distribution.
\subsection{Median buffer range percentage (BRP)}
The BRP metric \citep{Richards2011} corresponds to the fraction of fluxes whose deviation from the median is less than 10\% of the flux amplitude. This metric quantifies how concentrated the points are around the median values.
\subsection{$R_{cs}$}
The $R_{cs}$ metric \citep{kim2014} is the range of the cumulative sum of the normalized flux residuals. It is defined as
\begin{equation}
    R_{cs} = max(S_{l})-min(S_{l})
\end{equation}
Where $S_{l}$ is defined as:
\begin{equation}
    S_{l} = \frac{1}{n\sigma}\sum_{i}^{l}(F_{i}-\bar{F})
\end{equation}
With I=1,2,..,N. This metric characterizes the temporal coherence of the variability. Long-term coherent variations will produce large deviations from the mean, increasing $R_{cs}$. Constant fluctuations around the mean will cancel and lead to a smaller $R_{cs}$. This metric has been used for the classification of variable sources \citep{sanchez2021alerce}
\subsection{Smoothness index (s)}
The smoothness parameter s \citep{Ma2024} computes the second-order derivatives of the flux distribution to quantify the smoothness of a light curve. Similarly to $\eta$, large s values indicate rapid flux variations, whereas low s values indicate coherent continuous variations.
\begin{equation}
    s = \frac{1}{n} \sum \left|\frac{\delta F_{i+1}-\delta F{i}}{t_{i+2}-t_{i}}\right| \frac{\sqrt{\left(F_{i+2}-F_{i+1}\right)^{2} + \left(F_{i+1}-F_{i}\right)^{2}}}{\bar{F}}
    \label{eq:smooth}
\end{equation}
Where \textit{n} is the number of data points and $\delta F_{i}$ is defined as:
\begin{equation}
    \delta F_{i} = \frac{1}{\bar{F}} \frac{F_{i+1}-F_{i}}{t_{i+1}-t_{i}}
    \label{eq:smooth_delta}
\end{equation}
The value of s is strongly affected by the observational cadence \citep{Martinez2026}.
\subsection{Von Neumann ratio ($\eta$)}
The Von Neumann ratio \citep{von1941,von1942} quantifies the degree of correlation between consecutive flux measurements. It is similarly related to the smoothness of a light curve, helping to distinguish between coherent continuous variations and rapid uncorrelated behavior.
\begin{equation}
    \eta = \frac{\sum_{i}^{n-1} \frac{\left(F_{i+1}-F_{i}\right)^{2}}{n-1}}{\sum_{i}^{n} \frac{\left(F_{i}-\bar{F}\right)^{2}}{n-1}}
    \label{eq:eta}
\end{equation} 
A low value of $\eta$ is an indicator of highly correlated variability, whereas a high value indicates significant variations between consecutive points.
\subsection{Linear}
Slope of a linear best fit to the flux distribution \citep{Richards2011}. This metric may be used for characterizing long-term variability trends in the data and identifying changing state sources.
\subsection{Max slope}
The absolute value of the maximum slope between two consecutive observations \citep{Richards2011}. This metric is useful for identifying flare-like events.
\subsection{Pair slope}
The difference between the fraction of increasing flux changes and the fraction of decreasing flux changes. The flux changes are the difference between two consecutive photometric points. This metric is evaluated using the last 30 observations \citep{Richards2011}. A positive value indicates a flux increment in the last observations, whereas a negative value indicates a decrement. 
\subsection{Bayesian Block metrics}
\subsubsection*{$\Delta BB$}
Difference between the first and last fluxes in the Bayesian block representation of a time series \citep{graham2020,scargle2013}. 
\begin{equation}
    \Delta BB = |F_{init}^{BB}-F_{end}^{BB}|
\label{delta_BB}
\end{equation}
Similar to Linear, this metric can be used to identify objects that experience a change in state during the observed time.
\subsubsection*{BBA}
Amplitude of variation in the Bayesian block representation of a time series \citep{graham2020,scargle2013}.
\begin{equation}
    BBA = F_{max}^{BB}-F_{min}^{BB}
\label{BBA}
\end{equation}
This metric acts similarly to amplitude, but has the advantage of being less affected by noise and short-time non-physical fluctuations. 
\subsubsection*{Duty cycle ($\chi$)}
We define the Duty cycle ($\chi$) as the fraction of time spent in a flare-like state. We identify as a flare all the states in the Bayesian block representation above a certain threshold:
\begin{equation}
    F_{i} >  F_{q} + 3\sigma_{q}
    \label{eq:flare}
\end{equation}
Here $F_{q}$ represents the characteristic state of the object, which is calculated as the flux average of the Bayesian blocks, weighted by the time duration of each block, while $\sigma_{q}$ corresponds to the standard deviation, also weighted by the duration of the blocks. We identify $T_{flare}$ as the sum of all the times that match the condition of Eq. \ref{eq:flare}, whereas $T_{Total}$ is the total time covered by the Bayesian blocks. The Duty cycle ($\chi$) is then defined as:
\begin{equation}
    \chi = \frac{T_{flare}}{T_{Total}}
    \label{eq:duty}
\end{equation}
Our implementation of the duty cycle is inspired by multiple metrics used to identify flare events in Gamma-Ray blazars \citep{yoshida2023,ackermann2011}. A high value for this metric suggests flaring-like activity.
\subsubsection*{$\nu_{cp}$}
We define $\nu_{cp}$ as the ratio between the number of Bayesian blocks and the total observed time:
\begin{equation}
    \nu_{cp} = \frac{N_{blocks}}{T_{Total}}
\end{equation}
A high value of this metric indicates that an object exhibits complex variability, with multiple rapid fluctuations, whereas a low value indicates a smooth evolution.\\

Although we initially included the mean variance and Pair slope metrics, we excluded them from the final analysis. We found mean variance to be redundant given that the mean and standard deviation are already included in the set. On the other hand, the Pair slope has a discrete behavior, only adopting multiples of 1/29 between -1 and 1, complicating its interpretation in most of our analysis.

\newpage

\section{Metric distributions}
\label{app:metric_distributions}

\begin{table*}[h]
\centering
\caption{Statistical properties of the metric distributions. From left to right, columns show the average value (Mean), standard deviation (Std), 16th (P16), 50th (P50), and 84th (P84) percentiles, minimum and maximum values (Min, Max), third standardized moment of the distribution (Skewness), and the fourth standardized moment (Kurtosis). Metrics are calculated on g-band light curves.}
\label{tab:metric_statistics}
\small
\setlength{\tabcolsep}{6pt}
\begin{tabular}{lccccccccc}
\toprule
Metric & Mean & Std & P16 & P50 & P84 & Min & Max & Skewness & Kurtosis \\
\midrule
$\mu_{w}$ & 4.65e-27 & 6.72e-27 & 1.3e-27 & 2.8e-27 & 7.54e-27 & 8.46e-29 & 3.04e-25 & 13 & 392 \\
$A$ & 0.802 & 0.278 & 0.549 & 0.755 & 1.05 & 0.0836 & 4.43 & 1.35 & 5.15 \\
$\sigma_{w}$ & 0.1 & 0.0387 & 0.067 & 0.0917 & 0.134 & 0.0115 & 0.635 & 1.77 & 7.76 \\
$\mathrm{MAD}$ & 0.0768 & 0.0362 & 0.0497 & 0.0673 & 0.103 & 0.0107 & 0.551 & 2.96 & 16.6 \\
$\mathrm{RoMS}$ & 3.23 & 1.89 & 1.82 & 2.91 & 4.4 & 0.144 & 46.2 & 4.55 & 48.2 \\
$g_{1}$ & 1.61 & 0.67 & 1.01 & 1.58 & 2.24 & -2.28 & 5.32 & 0.169 & 1.23 \\
$g_{2}$ & 5.19 & 4.14 & 1.86 & 4.27 & 8.52 & -1.43 & 53.3 & 2.28 & 11.6 \\
$\mathrm{J}$ & 0.728 & 0.544 & 0.242 & 0.689 & 1.17 & -0.977 & 6.85 & 1.46 & 7.78 \\
$\mathrm{K}$ & 0.722 & 0.0392 & 0.687 & 0.719 & 0.757 & 0.389 & 0.911 & 0.456 & 1.93 \\
$F_{\mathrm{var}}$ & 0.107 & 0.0497 & 0.07 & 0.097 & 0.144 & 0.00573 & 1.93 & 6.38 & 156 \\
$\mathrm{s}$ & 0.00876 & 0.0194 & 0.00208 & 0.00523 & 0.0129 & 4.53e-05 & 1.22 & 27.5 & 1.34e+03 \\
$R_{cs}$ & 0.157 & 0.0824 & 0.0978 & 0.135 & 0.203 & 0.046 & 1.21 & 2.93 & 13.8 \\
$\mathrm{IQR}$ & 0.108 & 0.0545 & 0.0686 & 0.094 & 0.146 & 0.0145 & 1.14 & 3.7 & 31.5 \\
$\eta$ & 1.6 & 0.372 & 1.48 & 1.69 & 1.84 & 0.0105 & 2.33 & -2.5 & 6.53 \\
$\mathrm{BRP}$ & 0.68 & 0.119 & 0.574 & 0.694 & 0.792 & 0.134 & 0.96 & -0.883 & 1.05 \\
$\mathrm{Linear}$ & -5.28e-06 & 3.26e-05 & -2e-05 & -5.71e-06 & 9.78e-06 & -0.000478 & 0.000502 & 0.292 & 45.1 \\
$\mathrm{Max\ slope}$ & 4.38 & 15.6 & 0.319 & 1.11 & 4.72 & 0.0216 & 475 & 11.5 & 198 \\
$\mathrm{Mid20}$ & 0.113 & 0.0253 & 0.0913 & 0.109 & 0.132 & 0.0217 & 0.416 & 1.79 & 8.06 \\
$\mathrm{Mid35}$ & 0.206 & 0.0406 & 0.171 & 0.2 & 0.237 & 0.0408 & 0.636 & 1.54 & 5.6 \\
$\mathrm{Mid50}$ & 0.316 & 0.0545 & 0.269 & 0.308 & 0.36 & 0.0669 & 0.761 & 1.26 & 3.61 \\
$\mathrm{Mid65}$ & 0.463 & 0.0639 & 0.405 & 0.456 & 0.517 & 0.198 & 0.834 & 0.848 & 1.93 \\
$\mathrm{Mid80}$ & 0.698 & 0.0607 & 0.641 & 0.697 & 0.756 & 0.29 & 0.942 & 0.147 & 0.551 \\
$\Delta BB$ & 0.0949 & 0.117 & 0.0149 & 0.0576 & 0.166 & 0 & 2.08 & 3.5 & 23.2 \\
$BBA$ & 0.741 & 0.298 & 0.482 & 0.708 & 1.01 & 0 & 4.34 & 0.809 & 3.52 \\
$\chi$ & 0.0216 & 0.0147 & 0.0055 & 0.02 & 0.037 & 0 & 0.0968 & 0.482 & -0.13 \\
$\nu_{cp}$ & 0.0267 & 0.0187 & 0.00906 & 0.023 & 0.0443 & 0.000361 & 0.204 & 1.27 & 3.08 \\
\bottomrule
\end{tabular}
\end{table*}

\begin{figure*}[h]
    \centering
    \includegraphics[width=1.0\textwidth]{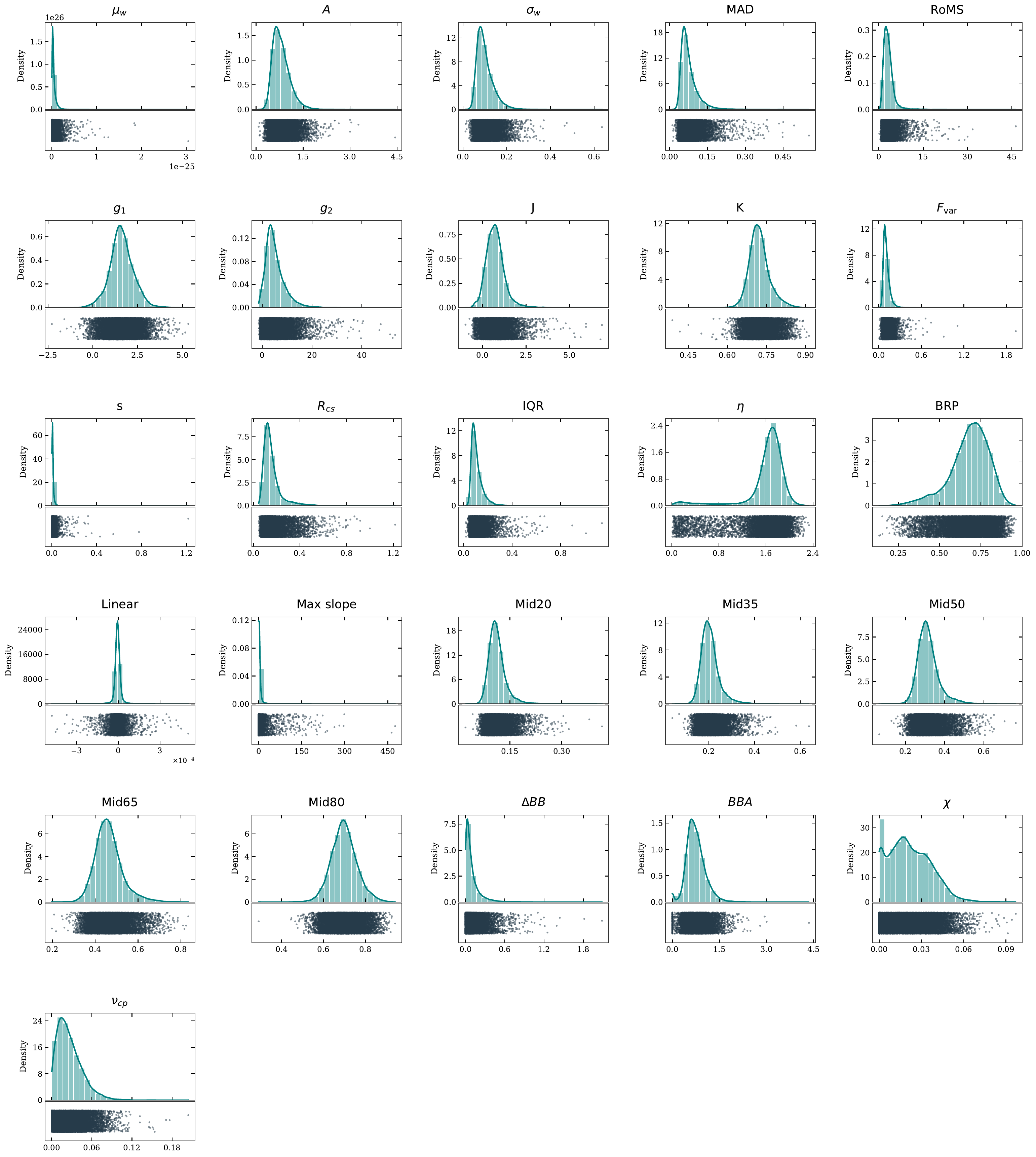}
    \caption{Complete set of metric distributions: The top of each panel shows the probability density histograms (bars) and the kernel density estimate (KDE, solid curve). The bottom shows the actual metric values for each source dispersed along the x-axis with a jitter for visualization. These distributions correspond to the $g$-band light curves.}
\label{fig:metric_dist_all}
\end{figure*}

\newpage

\section{Metrics catalog}
\label{app:catalog}
We provide the metric measurements for our sample in the form of a machine-readable table, which contains the following information.
\begin{table*}[ht]
\centering
\caption{Available measurements in the metric catalog. All metrics (except mean\_g and mean\_r) are calculated on the normalized light curves, i.e., divided by the average flux. Provided average fluxes, mean\_g and mean\_r, are in units of $\mathrm{erg}\,\mathrm{cm}^{-2}\,\mathrm{s}^{-1}\,\mathrm{Hz}^{-1}$. Object identifiers (Name), J2000 coordinates (RA, DEC), redshifts (Z), and classifications (Class) are adopted from the Million Quasars (Milliquas) catalog \citep{Flesch2023}. The original Milliquas spectral types were grouped into Type I, Type II, and Blazar (B) classes.}
\begin{tabular*}{\textwidth}{@{\extracolsep{\fill}}lr}
\hline
\textbf{Metric} & \textbf{Description} \\
\hline

Name & Object name. \\

RA & Right Ascension (J2000) in degrees. \\

DEC & Declination (J2000) in degrees. \\

Z & Redshift. \\

Class & AGN spectral classification. \\

mean\_g & Average weighted flux in the $g$-band. \\

mean\_r & Average weighted flux in the $r$-band. \\

Amp\_g & Flux amplitude in the $g$-band. \\

Amp\_r & Flux amplitude in the $r$-band. \\

std\_g & Flux-weighted standard deviation in the $g$-band. \\

std\_r & Flux-weighted standard deviation in the $r$-band. \\

MAD\_g & Median absolute deviation in the $g$-band. \\

MAD\_r & Median absolute deviation in the $r$-band. \\

ROMS\_g & Robust median statistic in the $g$-band. \\

ROMS\_r & Robust median statistic in the $r$-band. \\

Skewness\_g & Third standardized moment of the flux distribution in the $g$-band. \\

Skewness\_r & Third standardized moment of the flux distribution in the $r$-band. \\

Kurtosis\_g & Fourth standardized moment of the flux distribution in the $g$-band. \\

Kurtosis\_r & Fourth standardized moment of the flux distribution in the $r$-band. \\

J\_g & Stetson $J$ index in the $g$-band. \\

J\_r & Stetson $J$ index in the $r$-band. \\

K\_g & Stetson $K$ index in the $g$-band. \\

K\_r & Stetson $K$ index in the $r$-band. \\

Fvar\_g & Fractional variability amplitude in the $g$-band. \\

Fvar\_r & Fractional variability amplitude in the $r$-band. \\

s\_g & Smoothness index in the $g$-band. \\

s\_r & Smoothness index in the $r$-band. \\

R\_cs\_g & Range of the cumulative sum in the $g$-band. \\

R\_cs\_r & Range of the cumulative sum in the $r$-band. \\

\hline
\end{tabular*}
\label{tab:catalog_description_1}
\end{table*}

\addtocounter{table}{-1}
\begin{table*}[ht]
\centering
\caption{Available measurements in the metric catalog (continued).}
\begin{tabular*}{\textwidth}{@{\extracolsep{\fill}}lr}
\hline
\textbf{Metric} & \textbf{Description} \\
\hline

IQR\_g & Interquartile range in the $g$-band. \\

IQR\_r & Interquartile range in the $r$-band. \\

eta\_g & Von Neumann ratio in the $g$-band. \\

eta\_r & Von Neumann ratio in the $r$-band. \\

BRP\_g & Median buffer range percentage in the $g$-band. \\

BRP\_r & Median buffer range percentage in the $r$-band. \\

Linear\_g & Linear trend slope in the $g$-band. \\

Linear\_r & Linear trend slope in the $r$-band. \\

Max\_slope\_g & Maximum point-to-point slope in the $g$-band. \\

Max\_slope\_r & Maximum point-to-point slope in the $r$-band. \\

Mid20\_g & Central 20\% flux range normalized by the range $(P_{95}-P_{5})$ in the $g$-band. \\

Mid20\_r & Central 20\% flux range normalized by the range $(P_{95}-P_{5})$ in the $r$-band. \\

Mid35\_g & Central 35\% flux range normalized by the range $(P_{95}-P_{5})$ in the $g$-band. \\

Mid35\_r & Central 35\% flux range normalized by the range $(P_{95}-P_{5})$ in the $r$-band. \\

Mid50\_g & Central 50\% flux range normalized by the range $(P_{95}-P_{5})$ in the $g$-band. \\

Mid50\_r & Central 50\% flux range normalized by the range $(P_{95}-P_{5})$ in the $r$-band. \\

Mid65\_g & Central 65\% flux range normalized by the range $(P_{95}-P_{5})$ in the $g$-band. \\

Mid65\_r & Central 65\% flux range normalized by the range $(P_{95}-P_{5})$ in the $r$-band. \\

Mid80\_g & Central 80\% flux range normalized by the range $(P_{95}-P_{5})$ in the $g$-band. \\

Mid80\_r & Central 80\% flux range normalized by the range $(P_{95}-P_{5})$ in the $r$-band. \\

BB\_delta\_g & Difference between the first and last flux in the Bayesian block representation of the $g$-band light curve. \\

BB\_delta\_r & Difference between the first and last flux in the Bayesian block representation of the $r$-band light curve. \\

BB\_amp\_g & Flux amplitude in the Bayesian block representation of the $g$-band light curve. \\

BB\_amp\_r & Flux amplitude in the Bayesian block representation of the $r$-band light curve. \\

Duty\_cycle\_g & Fraction of time spent in an active state in the $g$-band. \\

Duty\_cycle\_r & Fraction of time spent in an active state in the $r$-band. \\

nu\_cp\_g & Number of Bayesian blocks normalized by the time baseline in the $g$-band. \\

nu\_cp\_r & Number of Bayesian blocks normalized by the time baseline in the $r$-band. \\

\hline
\end{tabular*}
\label{tab:catalog_description_2}
\end{table*}

\end{document}